\documentstyle[12pt,aaspp4,flushrt]{article}
\begin{document}
\renewcommand{\topfraction}{1}
\renewcommand{\bottomfraction}{1}
\renewcommand{\textfraction}{0}
\renewcommand{\floatpagefraction}{2}
\textfloatsep=15pt
\intextsep=15pt
\floatsep=15pt
\renewcommand{\bottomfraction}{1}

\def\gtorder{\mathrel{\raise.3ex\hbox{$>$}\mkern-14mu
             \lower0.6ex\hbox{$\sim$}}}
\def\ltorder{\mathrel{\raise.3ex\hbox{$<$}\mkern-14mu
             \lower0.6ex\hbox{$\sim$}}}

\centerline{\Large Large-Scale Structure in the Universe}
\vglue30pt

\vglue-15pt
\centerline{\large\textsc{Neta A. Bahcall}}
\vglue15pt
\centerline{Princeton University Observatory, Princeton, NJ}

\vglue30pt

\section*{\textsc{Abstract}}

How is the universe organized on large scales?  How did this
structure evolve from the unknown initial conditions to the present
time? 
 The answers to these questions will shed light on the cosmology we
live in,
 the amount, composition and distribution of matter in the universe,
the initial
 spectrum of density fluctuations, and the formation and evolution of
galaxies, clusters
 of galaxies, and larger scale structures.

I review observational studies of
  large-scale structure,  describe
  the peculiar velocity field on large scales,
and present 
 the observational constraints placed on  cosmological models and
 the mass density of the universe.  While major progress has
been made
over the last decade, most observational problems in the field of large-scale
structure remain open.  I will highlight some of the
unsolved
 problems that seem likely to be solved in the
 next decade.

\section{\textsc{Introduction}}
\label{sec:1}

	The existence of large-scale structure in the
 universe has been known for over half a century (e.g.,
\cite{shapley30,zwicky57}).  A spectacular increase in our
understanding of this subject
 has developed over the last decade, led by observations of the
distribution of galaxies and of clusters of galaxies.  With
 major surveys underway, the next decade will provide  new
milestones in
 the study of large-scale structure.  I will
highlight
 what we currently know about large-scale structure, emphasizing some
of the
 unsolved problems and what we can hope to learn in the next
 ten years.

Why study large-scale structure? We all want to know what the
``skeleton'' of our universe
 is, that is, how does the universe look on the largest scales we can
investigate?   It is natural to ask: on what scales does the universe
become homogeneous and isotropic, as expected from the cosmological
principle? In addition, we
shall see that detailed
 knowledge of the large-scale structure provides constraints on:  
\begin{itemize}
\item the formation and evolution of galaxies and larger structures;
\item the
 cosmological model of our universe (including the mass density of the
universe, the
 nature and amount of the dark matter, and the initial spectrum of fluctuations
that
 gave rise to the structures seen today).
\end{itemize}

What have we learned so
 far, and what are the main unsolved problems in the field of
 large-scale structure?  I discuss these questions in the sections that follow.
I first list some of the unsolved problems. 

 There are many fundamental problems on which progress is likely to be
made in the next decade.  In fact, as we go through what is known
about large-scale structure, you will see that most of what we want to
know is a challenge for the future.
Here is a partial list of some of the most
 interesting unsolved problems.
\begin{itemize}
\item Quantify the measures of large-scale structure.  
How large are the largest coherent structures?  How strong is the clustering
on
 large scales  (e.g., as quantified by the power spectrum and the correlation
functions of
 galaxies and other systems)?
\item What is the topology of large-scale structure?  What are the 
shapes and morphologies of superclusters, voids,
filaments, and their
 networks?
\item How does large-scale structure depend on galaxy type, luminosity,
surface
 brightness? How does the large-scale distribution of galaxies differ
from that of other systems (e.g., clusters, quasars)?
\item What is the amplitude of the peculiar velocity field as a
function of scale? 
\item What is the amount of mass and
 the distribution of mass on large scales?
\item Does mass trace light on large scales?  What is in the ``voids"?
\item What is the mass density, $\Omega_m \equiv \rho_m/
\rho_{\mathrm{crit}}$, of the
 universe?
\item What is the baryon fraction of the universe, $\Omega_b/\Omega_m$?
\item How does the large-scale structure evolve with time?
\item What
 are the implications of the observed large-scale structure for
 the cosmological model of our universe and
 for structure formation? (e.g., What is the nature of the dark
matter?  Does structure form by gravitational instability?  What is
the initial spectrum of fluctuations that gave rise to the structure
we see today?  Were the fluctuations Gaussian?)
\end{itemize}

\begin{figure}[t]
\plotfiddle{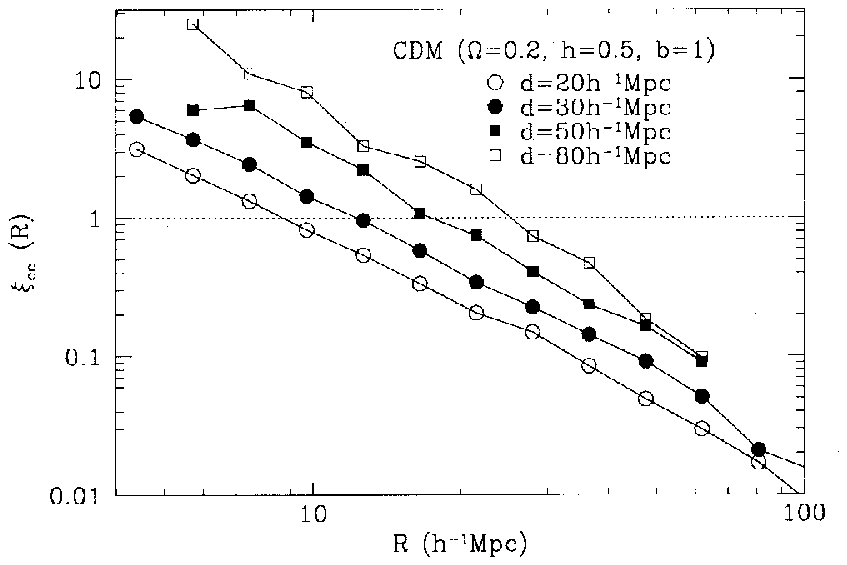}{12cm}{0}{200}{200}{-234}{72}
\caption[]{  Redshift cone diagram for 
galaxies in the Las Campanas survey
\protect\cite{kirshner95}\label{fig:one}.}
\end{figure}

\begin{figure}[ht]
\plotfiddle{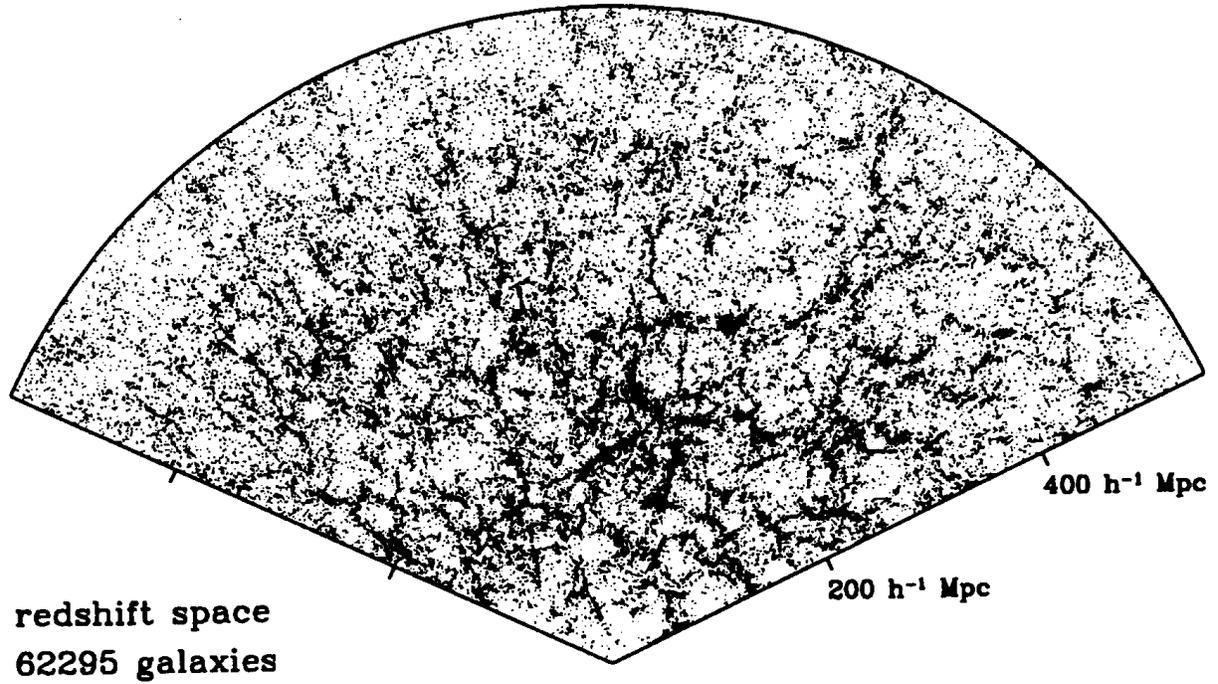}{12cm}{0}{110}{110}{-234}{72}
\caption[]{  The redshift-space distribution
 of galaxies in a $6^\circ$ thick slice along the SDSS survey equator
 from a large N-body cosmological simulation (cold-dark-matter with 
$\Omega_m h \simeq 0.24$; Gott, et al., in preparation).  This slice 
contains approximately 6\% of the $10^6$ 
 galaxies in the spectroscopic survey\label{fig:two}.}
\end{figure}

Two-dimensional surveys of the universe
 analyzed with 
correlation function statistics \cite{groth77,maddox90} reveal 
structure to
 scales of at least $\sim 20 h^{-1}$~Mpc.  Large redshift surveys
of the
 galaxy distribution reveal a considerably more detailed structure
of superclusters, voids, and filament
 network extending to scales of $\sim 50$--$100h^{-1}$~Mpc 
(\cite{gregory78,gregory81,chin81,giov86,delap86,dacosta88,geller89}).  The
 most recent and largest redshift survey, the Las Campanas Redshift
Survey  (\cite{kirshner95}; 
see also \cite{landy96}), is
presented in Figure~\ref{fig:one}; it reveals the ``cellular" nature of the 
large-scale galaxy distribution.
The
 upcoming Sloan Digital Sky Survey (SDSS), expected to begin operation
in 1997 
 (see \S~\ref{sec:6}), will provide a three dimensional map of the entire 
high-latitude
 northern sky to $z \sim 0.2$, with redshifts for approximately $10^6$
galaxies.  
Figure~\ref{fig:two} presents a simulation of one expected redshift slice from
the
 SDSS survey; the slice represents $\sim 6\%$ of the total survey.  This
survey, and others currently planned, will provide the large increase in the
survey volume
  required to resolve some of the unsolved problems
listed above (see \S~\ref{sec:6} for details).

\begin{figure}[ht]
\plotfiddle{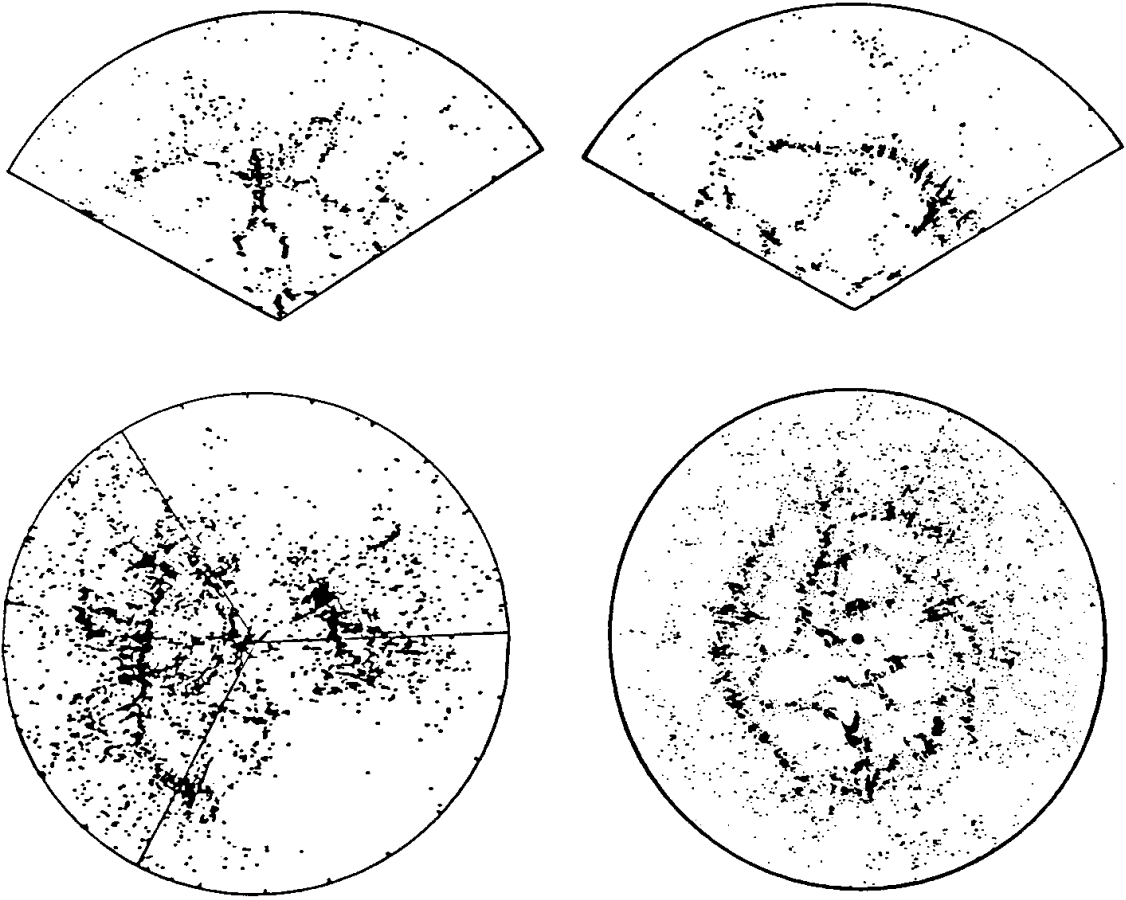}{12cm}{0}{125}{125}{-234}{72}
\caption[]{  Redshift cone diagrams for 
observed (CfA survey) and simulated
 (CDM cosmology with $\Omega_m h = 0.3$) galaxy (mass) 
distributions in the present universe.
Which are data
 and which are simulations?  The similarity of the two indicates the
underlying
 success of basic cosmology in reproducing the observed 
structure\label{fig:three}.}
\end{figure}

One of the exciting new developments in
 the last decade is the ability to compare the large-scale
observations
 with expectations from different cosmologies using state-of-the-art
numerical cosmological
 simulations [see the contribution by J.~P.~Ostriker in these
proceedings]. 
 Standard  cosmological models
generally succeed in
 reproducing the overall large-scale structure that we observe.  In
Figure~\ref{fig:three}, I present two examples of the 
observed versus simulated redshift
cone diagrams
 (the latter for a cold-dark-matter, CDM, model).  I leave it
 up to the viewer to decide which figure displays data and which
 displays a simulation, and how well they match.  The ``good" news is
 that a simple cosmological model, starting from a rather uniform
distribution of
 matter after the big-bang, with minute density fluctuations as
prescribed by
 a CDM power spectrum [see the presentations by J. P. Ostriker, P. J.
E. Peebles, and P. Steinhardt], can represent the universe we
see
 today reasonably well.  This is a real success for cosmology!  The ``bad" 
news is
 that the quantitative match with the simplest standard models is not
very
 good.  Model parameters need to be adjusted, some in an ad-hoc
 manner, in order to fit the data quantitatively.  

\section{\textsc{Clustering and Large-Scale Structure}}
\label{sec:2}

I summarize in Section \ref{sec:2.1} the ways that individual galaxies reveal
the large-scale structure of the universe and describe in 
Section \ref{sec:2.2}
the study of large-scale structure using clusters of galaxies.

\subsection{\textit{Galaxies and Large-Scale Structure}}
\label{sec:2.1}

The correlation function (of galaxies, of clusters of galaxies, of
quasars) 
 is a\break powerful
 statistical measure of the clustering strength on different scales.
The two-point
 spatial correlation function, $\xi (r)$, is the net probability above random 
of
finding
 a pair of objects, each within a given volume element $dV_i$, separated
 by the distance $r$.  The total probability of finding such a pair
 is given by:  
\begin{equation}
dP = n^2 (1+ \xi (r)) dV_1 dV_2\ , 
\label{eq:one}
\end{equation}
where $n$
 is the mean density of objects in the sample.

\begin{figure}[ht]
\plotfiddle{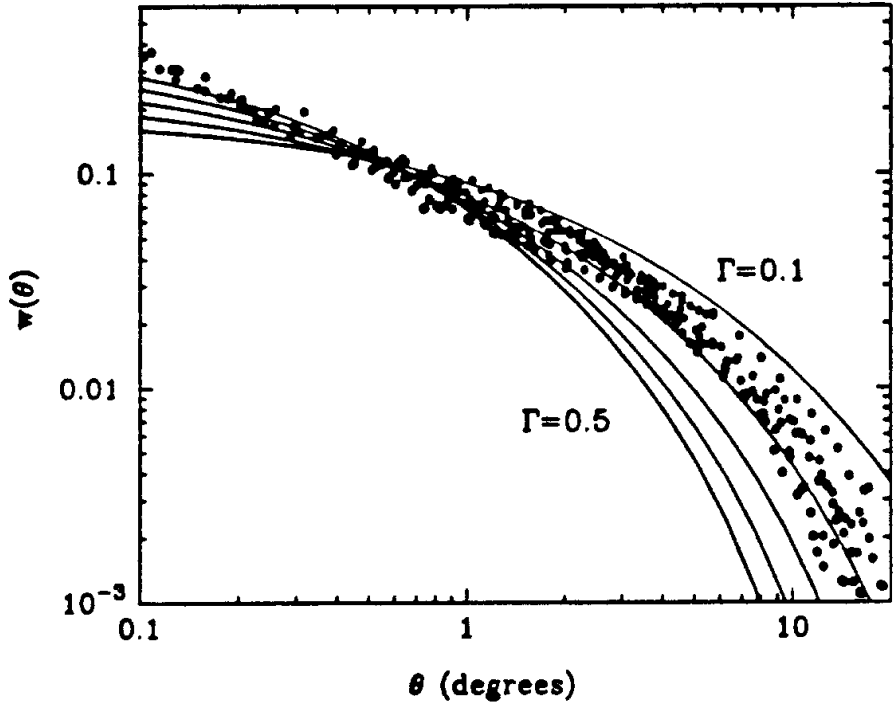}{12cm}{0}{125}{125}{-198}{72}
\caption[]{  The
 scaled angular correlation function of galaxies measured from the APM
survey plotted
 against linear theory predictions for CDM models (normalized to
$\sigma_8 = 1$ on $8 h^{-1}~{\mathrm{Mpc}}$ scale) with
$\Gamma \equiv \Omega_m h = 0.5, 0.4, 0.3, 0.2$ and 0.1
\protect\cite{efs90}\label{fig:four}.}
\end{figure}

The angular (projected) galaxy correlation function
 was first determined from the 2D Lick survey and inverted into a
 spatial correlation function by Groth \& Peebles \cite{groth77}.  They
find $\xi_{gg}(r) \simeq 20 r^{-1.8}$ for $r \ltorder 15 h^{-1}$~Mpc, with 
correlations that
 drop to the level of the noise for larger scales.  This observation implies 
that
galaxies
 are clustered on at least  $\ltorder 15 h^{-1}$~Mpc scale, with a 
 correlation scale of $r_o(gg) \simeq 5 h^{-1}$~Mpc, where $\xi (r) 
\equiv (r/r_o)^{-1.8} \equiv Ar^{-1.8}$.  More recent results 
support the above conclusions
(with
 possibly a weak tail to larger scales in the galaxy correlations).  One of the
 recent determinations of the two-point angular galaxy correlation
function, using the
 APM 2D galaxy survey \cite{maddox90,efs90}, is presented in 
Figure~\ref{fig:four}.  The
 observed correlation function is compared with expectations from the
CDM cosmology (using
 linear theory estimates) for different values 
of the 
parameter $\Gamma =\Omega_m h$.  Here 
$\Omega_m$ is the mass density of the universe in
terms of the critical density and  $h \equiv H_0/100~
\mathrm{km~ s^{-1}~
Mpc^{-1}}$ [see the contribution of Peebles].
The different $\Omega_m h$ models differ mainly in
 the large-scale tail of the galaxy correlations:  higher values of
$\Omega_m h$
 predict less structure on large scales (for a given normalization of
the initial mass fluctuation spectrum) since the CDM fluctuation
spectrum (see below) 
peaks
 on scales that are inversely proportional to $\Omega_m h$.  It is clear from
 Figure~\ref{fig:four}, as was first shown from the analysis of galaxy 
clusters (see below),
that
 the standard CDM model with $\Omega_m = 1$ and $h = 0.5$ does
 not produce enough large-scale power to match the observations.
As Figure~\ref{fig:four} shows, the galaxy correlation function
requires $\Omega_m h \sim 0.15$--0.2 for a CDM-type spectrum, consistent with
 other large-scale structure data.

\begin{figure}[ht]
\plotfiddle{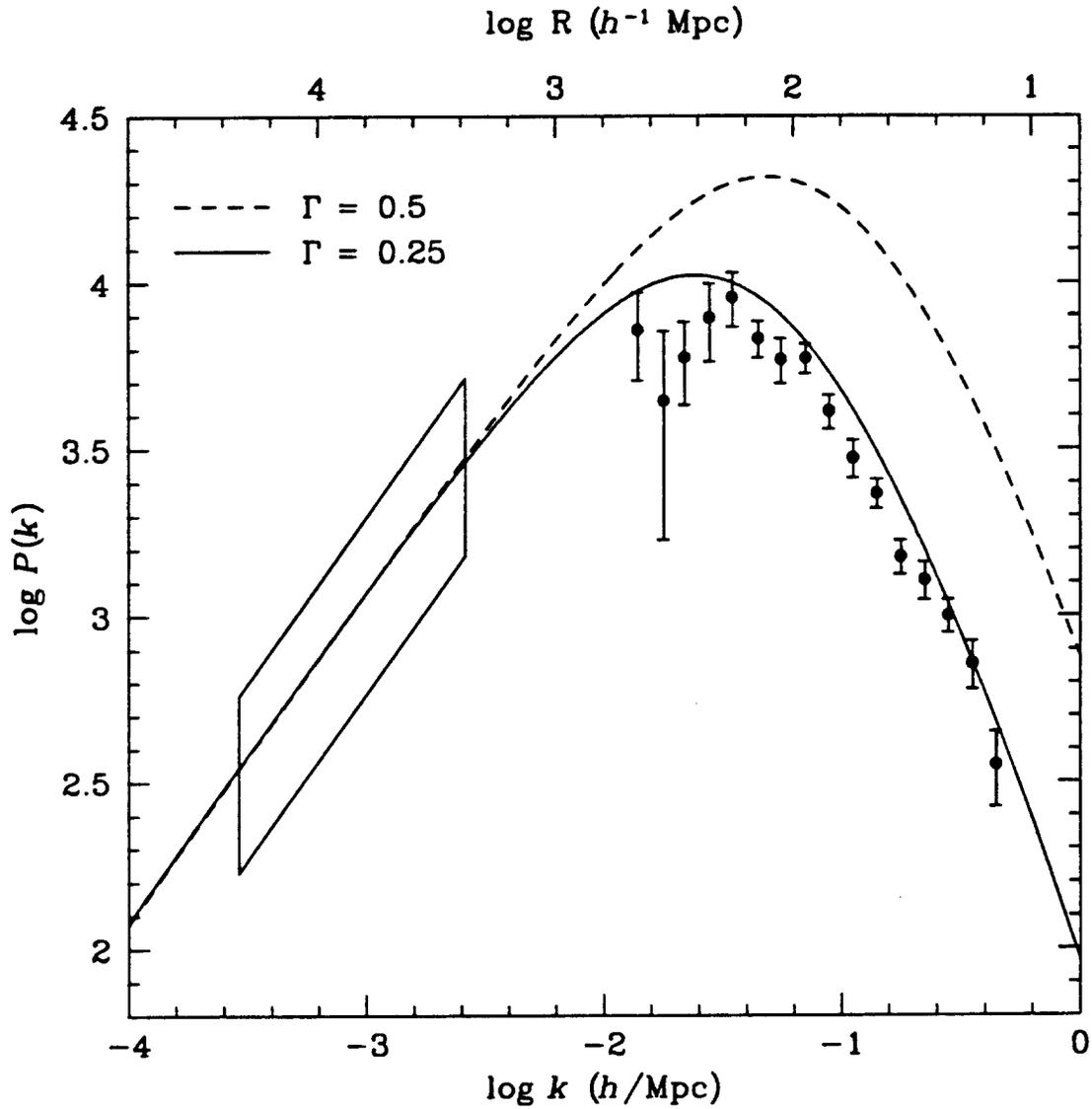}{12cm}{0}{100}{100}{-234}{36}
\caption[]{  The power spectrum as derived from a variety of
 tracers and redshift surveys, after correction for non-linear effects, 
redshift
distortions, and relative
 biases; from \cite{peacock94}.  The two curves show the
Standard CDM
 power spectrum ($\Gamma = 0.5$), and that of CDM with $\Gamma = 0.25$.  Both 
are
 normalized to the COBE fluctuations, shown as the box on the
left-hand side of the figure\label{fig:five}.}
\end{figure}

The power spectrum, $P(k)$, which reflects the initial spectrum
of fluctuations [see the discussion by P. Steinhardt in these
proceedings] 
that gave rise to
 galaxies and other structure, is
represented by the Fourier transform of the
 correlation function
(e.g., \cite{peebles93}).  One of the recent attempts to determine 
this fundamental statistic using a variety of tracers is presented 
in Figure~\ref{fig:five} (\cite{peacock94}; see
also \cite{vogeley92,fisher93,park94,landy96}).  The determination of this 
composite
spectrum assumes different
 normalizations for the different tracers used (optical galaxies, IR
galaxies, clusters of
 galaxies).  The different normalizations imply a different bias
parameter $b$ for each of the different tracers [where $b \equiv
(\Delta \rho/\rho)_{\mathrm{gal}}/(\Delta \rho/\rho)_m$ represents the
overdensity of the galaxy tracer relative to the mass overdensity]. 
Figure~\ref{fig:five} also shows the microwave background radiation
(MBR) anisotropy as
 measured by COBE \cite{smoot92} on the largest scales 
($\sim 1000 h^{-1}$~Mpc) and compares the data with the mass power
spectrum expected for two CDM models:  a
standard
 CDM model with $\Omega_m h = 0.5\  (\Omega_m = 1, h = 0.5$), and
 a low-density CDM model with $\Omega_m h = 0.25$.  The latter model appears
 to provide the best fit to the data, given the normalizations used
 by the authors for the different galaxy tracers.

\begin{figure}[ht]

\plotfiddle{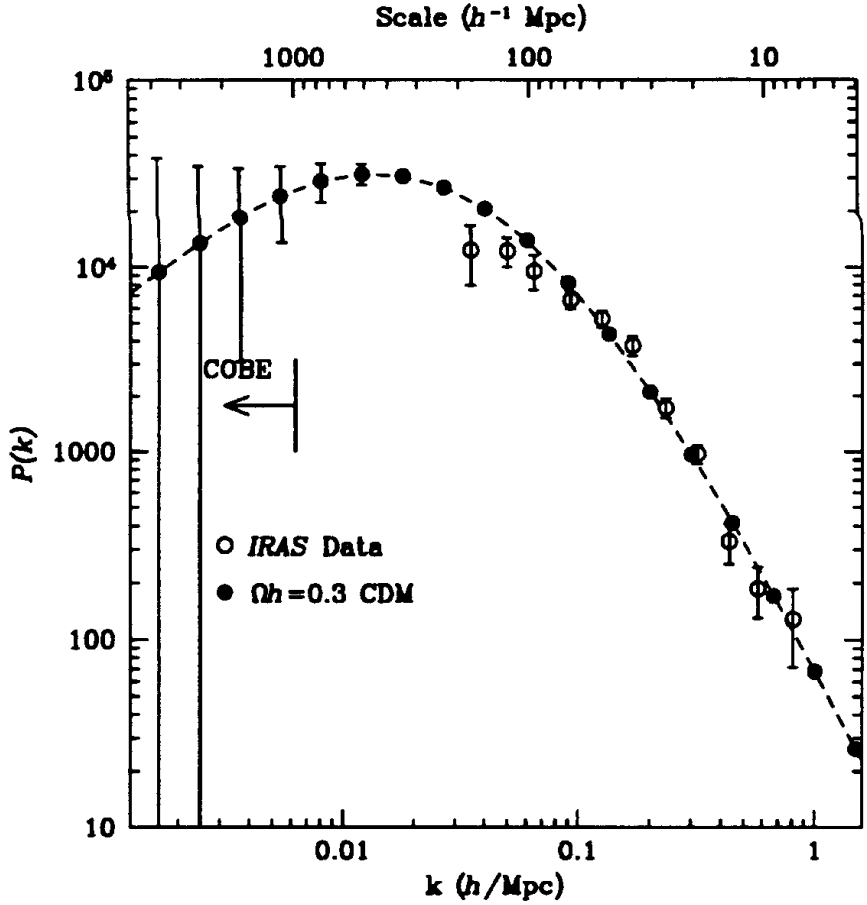}{12cm}{0}{120}{120}{-198}{36}
\caption[]{  Estimate of the galaxy power
 spectrum that can be obtained with the Sloan Survey.  The solid
points
 and dashed line show the linear theory power spectrum of 
$\Omega_m h = 0.3$~CDM, with error bar estimates appropriate for the Sloan 
Survey.
Open
 points are from \cite{fisher93}.  The scales probed by COBE
 are shown\label{fig:six}.}
\end{figure}

The next decade will provide critical advances in the
 determination of the\break correlation function and the related power 
spectrum.
The large redshift
 surveys now underway (\S~\ref{sec:6}) will probe the power spectrum of 
galaxies
to
 larger scales than currently available and with greater
accuracy.  These surveys
 will bridge the gap between the current optical determinations of 
$P(k)$ of galaxies on scales\break $\ltorder 100 h^{-1}$~Mpc and the MBR 
anisotropy on 
scales $\gtorder 10^3 h^{-1}$~Mpc.  This bridge will cover the 
 critical range of the
 spectrum turnover, which reflects the horizon scale at the time of
matter-radiation equality.  An example of what may be expected from the
Sloan
 survey is shown in Figure~\ref{fig:six}; a clear determination of $P(k)$ over 
a
wide 
range of scales is expected in the near future. This will enable the
determination of the initial spectrum of fluctuations at recombination
that gave rise to the structure we see today and will shed light on the
cosmological model parameters that may be responsible for that
spectrum (such as $\Omega_m h$  and the nature of the dark matter). 
 In the next decade,  
 $P(k)$ will also be determined from the MBR anisotropy measurements on small 
scales ($\sim 0.1^\circ$
 to  $\sim 5^\circ$) [see P. Steinhardt's contribution in these
proceedings],  allowing a most important
overlap in the determination of the galaxy $P(k)$ from redshift
surveys and the mass $P(k)$ from the MBR anisotropy.  
These data will place  constraints
on cosmological parameters including $\Omega (= \Omega_m +
\Omega_\Lambda), \Omega_m,\
\Omega_b, h$, and the nature of the dark matter itself.

\subsection{\textit{Clusters and Large-Scale Structure}}
\label{sec:2.2}

The correlation function of clusters of galaxies 
  efficiently  quantifies the large-scale structure of the
universe.  Clusters are correlated in space more strongly than 
are individual galaxies, by
an order of
 magnitude, and their correlation extends to considerably larger
scales ($\sim 50 h^{-1}$~Mpc).  The cluster correlation strength 
increases with richness ($\propto$ luminosity or mass) of the
 system from single galaxies to the richest clusters
\cite{bahcall83,bahcall88}.  
Here ``richness" refers to the number of cluster
galaxies within
 a given linear radius of the cluster center and within a given
 luminosity range (e.g., \cite{abell58}).  The correlation strength also
increases with
 the mean spatial separation of the clusters
\cite{szalay85,bahcall86}.  
This dependence results in a ``universal"
dimensionless cluster correlation
 function; the cluster dimensionless correlation scale is constant for
all clusters when
 normalized by the mean cluster separation.

Empirically, two general relations have been found  \cite{bahcall92a}
for the correlation\break function of clusters of galaxies, $\xi_i = A_i
r^{-1.8}$: 
\begin{equation}  
A_i \propto N_i\ ,
\label{eq:two}
\end{equation}
\begin{equation}
A_i \simeq (0.4 d_i)^{1.8}\ ,
\label{eq:three}
\end{equation}
 where $A_i$ is
 the amplitude of the cluster correlation function, 
$N_i$ is the richness 
of
 the galaxy clusters of type $i$, and $d_i$ is the mean separation of the
 clusters.  Here $d_i = n_i^{-1/3}$, where $n_i$ is the mean spatial
 number density of clusters of richness $N_i$ in a volume-limited,
richness-limited complete sample.  The first relation, equation~(\ref{eq:two}),
states that the amplitude of the cluster correlation function
increases with cluster richness, i.e., rich clusters are more strongly
correlated than poorer clusters.  The second relation,
equation~(\ref{eq:three}), 
states
that the amplitude of the cluster correlation function depends on the
mean separation of clusters (or, equivalently, on their number
density); the rarer, large mean separation richer clusters are more
strongly correlated than the more numerous poorer clusters. 
 Equations~(\ref{eq:two}) and (\ref{eq:three}) relate to 
each other through the richness function
of clusters, i.e., the number density of clusters as a function of
their richness.  Equation (\ref{eq:three}) 
 describes a universal scale-invariant (dimensionless) correlation
function with  a correlation
 scale $r_{o,i} = A_i^{1/1.8} \simeq 0.4 d_i$ (for $30 \ltorder d_i
\ltorder 90 h^{-1}~{\mathrm{Mpc}}$).

There are some conflicting statements in the literature about the
precise values of the correlation amplitude, $A_i$.  Nearly all these
contradictions are caused by not taking account of equation~(\ref{eq:two}). 
 When apples are
 compared to oranges, or the clustering of rich clusters is compared
to
 the clustering of poorer clusters, differences are expected and
observed.

\begin{figure}[ht]
\plotfiddle{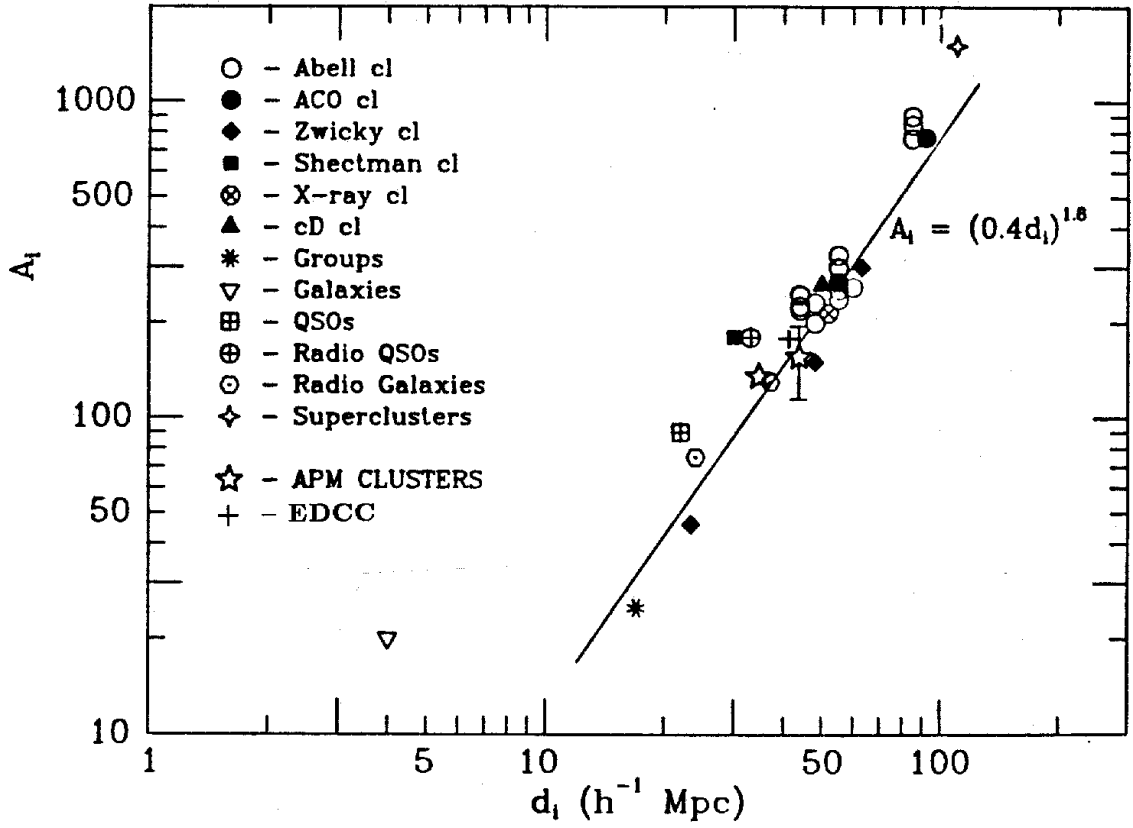}{12cm}{0}{120}{120}{-250}{36}
\caption[]{  The dependence of the cluster correlation amplitude on
 mean cluster separation \cite{bahcall92a}\label{fig:seven}.}
\end{figure}

Figure~\ref{fig:seven} clarifies the observational situation.  
The $A_i(d_i)$ relation for groups and clusters of various
richnesses is
 presented in the figure.  The recent automated cluster surveys of APM
\cite{dalton92} and EDCC \cite{nichol92} are 
consistent with the predictions of equations~(\ref{eq:two}) and
(\ref{eq:three}), 
as is the
correlation
 function of X-ray selected ROSAT clusters of galaxies
\cite{romer94,bahcall94a}. 
Bahcall and Cen \cite{bahcall94a}
  show that a flux-limited sample of
 X-ray selected clusters will exhibit a correlation scale that is
smaller
 than that of a volume-limited, richness-limited sample of comparable
apparent
 spatial density since the flux-limited sample contains poor groups nearby
and only
 the richest clusters farther away.  Using the richness-dependent
cluster correlations of equations~(\ref{eq:two}) and (\ref{eq:three}), 
Bahcall and Cen \cite{bahcall94a} find excellent agreement with the
observed flux-limited
 X-ray cluster correlations of Romer et al.~\cite{romer94}.  Comparison of the
 observed cluster correlation function with cosmological models
strongly constrains the model parameters
 (see below).

\begin{figure}[ht]
\plotfiddle{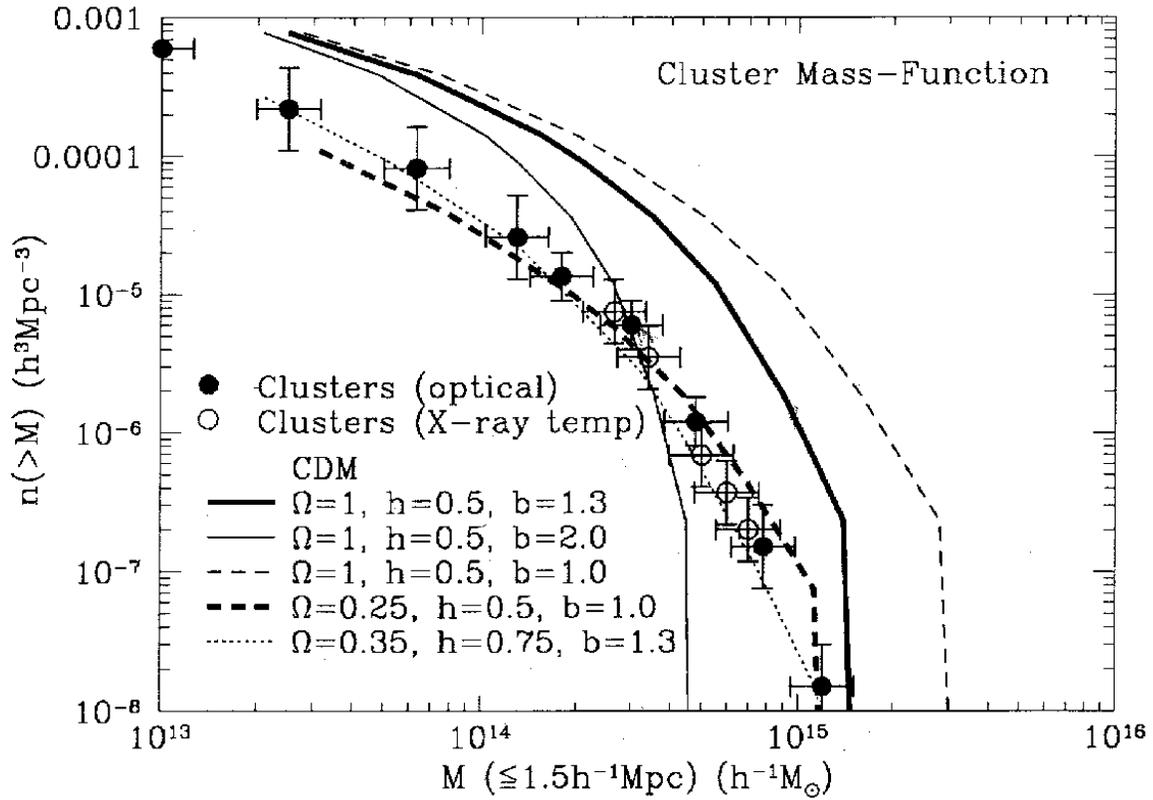}{12cm}{0}{160}{160}{-250}{36}

\caption[]{  The mass function
 of clusters of galaxies from observations (points) and cosmological
simulations of different
 $\Omega_m h$ CDM models \cite{bahcall92b,bahcall93}\label{fig:eight}.}
\end{figure}

The observed mass function (MF), $n(>M)$, of clusters of galaxies,
which describes  
the
 number density of clusters above a threshold mass $M$,  
 can be used as a critical test of theories of structure formation in
the universe.  The richest, most massive clusters are thought to 
form from rare high peaks in the
initial
 mass-density fluctuations; poorer clusters and groups form from
smaller, more common
 fluctuations.  
Bahcall and Cen \cite{bahcall93} determined the MF of clusters 
of galaxies using both optical
and
 X-ray observations of clusters.  Their MF is presented in 
Figure~\ref{fig:eight}.  
The function is well fit by the analytic expression
\begin{equation} 
n(>M) = 4 \times 10^{-5} (M/M^*)^{-1} \exp (- M/M^*) h^3~
{\mathrm{Mpc^{-3}}}\ ,
\label{eq:four}
\end{equation} 
with $M^* = (1.8 \pm 0.3) \times 10^{14} h^{-1}~ M_\odot$, (where the mass
 $M$ represents the cluster mass within $1.5 h^{-1}$~Mpc radius).

\begin{figure}[ht]
\plotfiddle{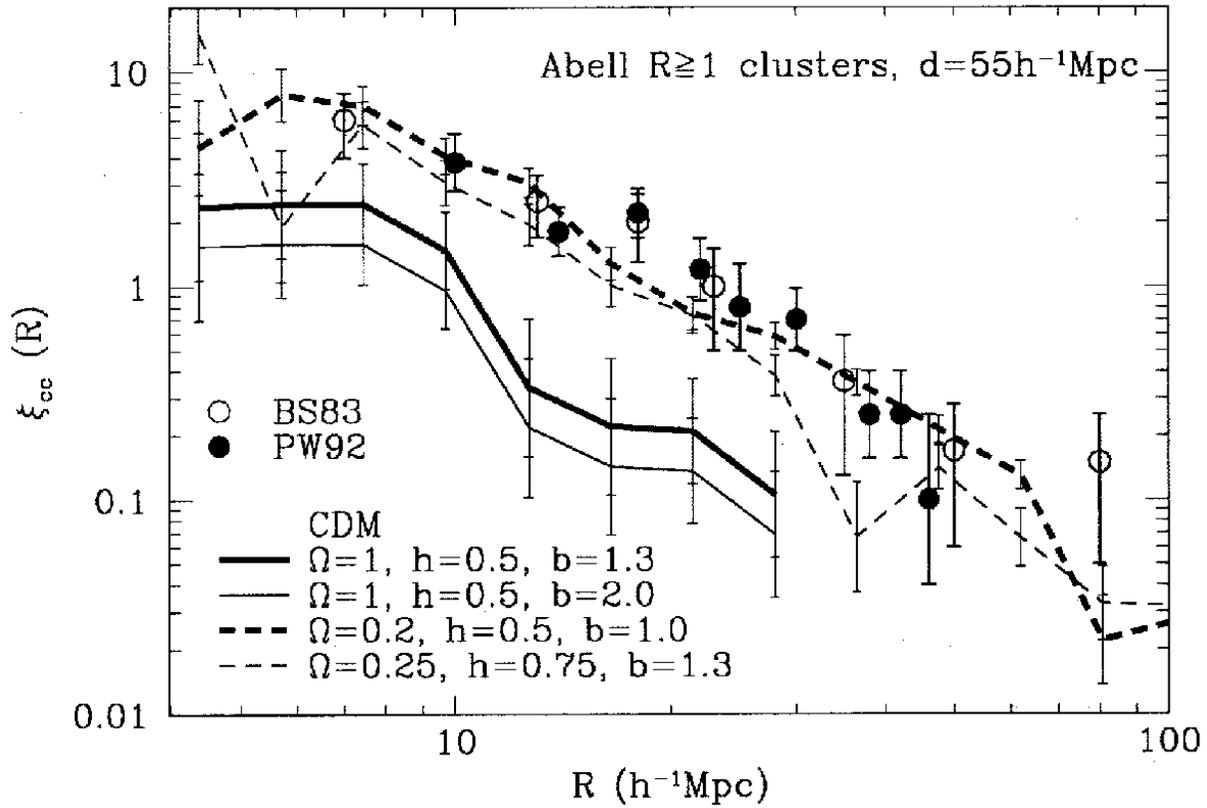}{12cm}{0}{160}{160}{-250}{36}

\caption[]{  The correlation
 function of rich ($R \geq 1$) Abell clusters, with $d~= 55
h^{-1}$~Mpc, 
from observations \cite{bahcall83,peacock92} and 
simulations \cite{bahcall92b}\label{fig:nine}.}
\end{figure}

\begin{figure}[ht]
\plotfiddle{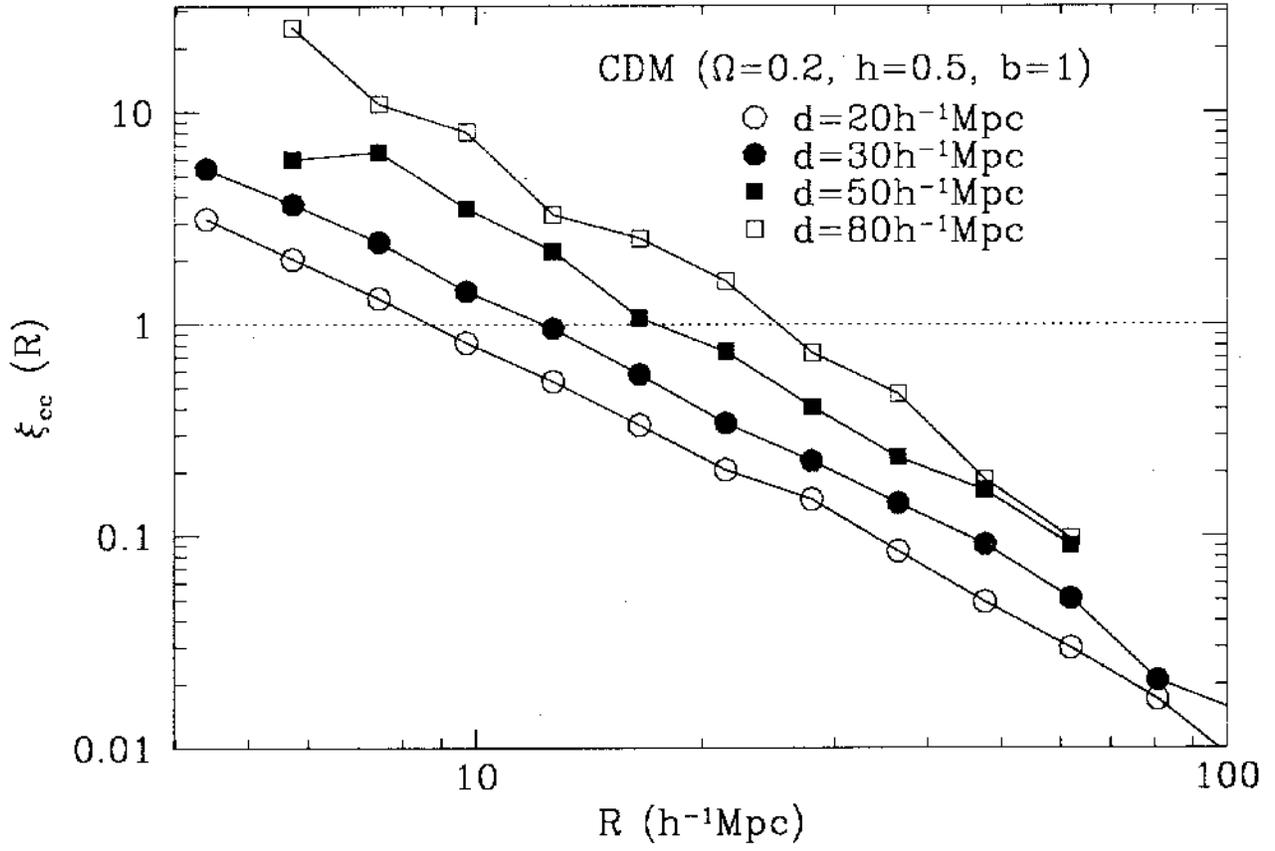}{12cm}{0}{200}{200}{-250}{36}

\caption[]{  Dependence of
 the model cluster correlation function on mean cluster separation
$d$\label{fig:ten}.}
\end{figure}

\begin{figure}[ht]
\plotfiddle{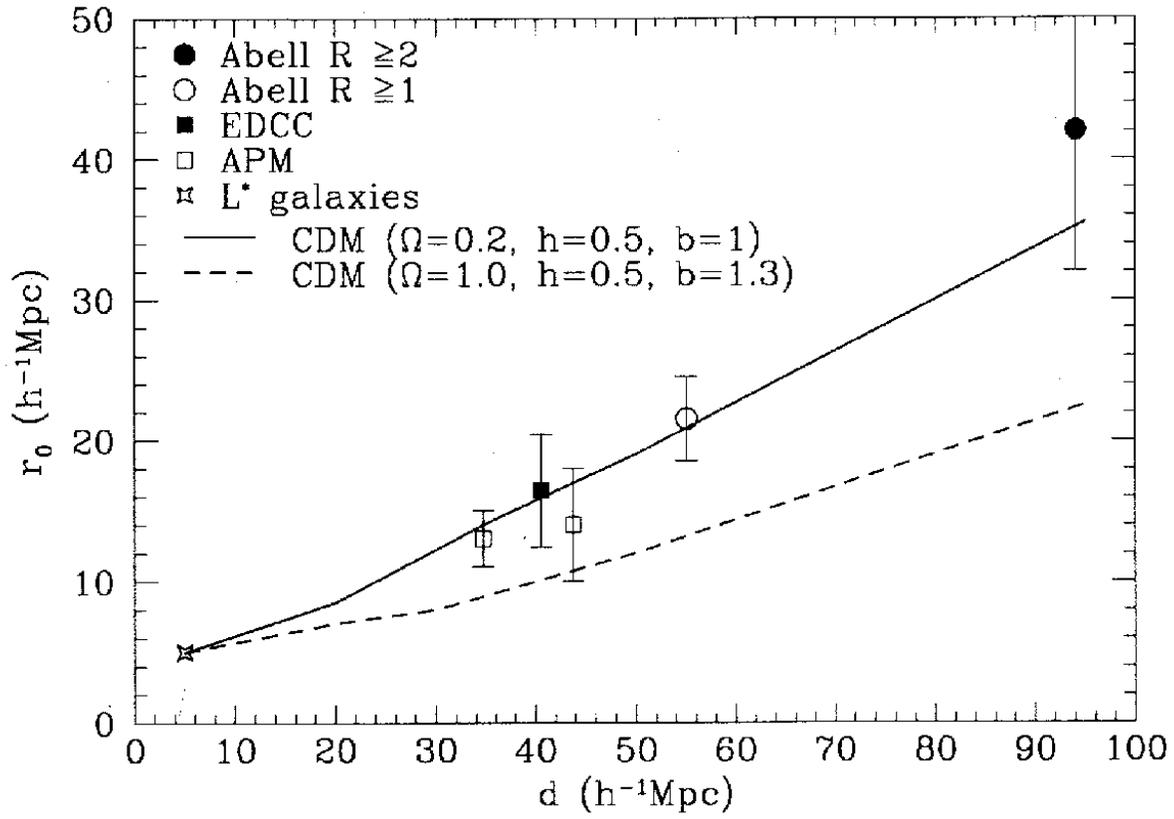}{12cm}{0}{160}{160}{-250}{36}
\caption[]{  The correlation length of 
the cluster correlation function as
a function of
 mean cluster separation, from observations and cosmological
simulations \cite{bahcall92b}\label{fig:eleven}.}
\end{figure}

Bahcall and Cen \cite{bahcall92b} compared 
the observed mass function and correlation\break function
 of galaxy clusters with predictions of N-body cosmological simulations
of
 standard $(\Omega_m = 1)$ and
 nonstandard $(\Omega_m < 1)$ CDM models. They find 
that none of the standard $\Omega_m = 1$~CDM 
 models, with any normalization, can reproduce both the observed 
correlation function and the mass
function of clusters.  A
low-density
 ($\Omega_m \sim 0.2$--0.3) CDM-type model, however, provides a good
 fit to both sets of observations (see Figs.~\ref{fig:eight}--\ref{fig:eleven})
.  
 The  constraints on $\Omega_m$ 
 are model dependent; a mixed hot + cold dark matter model, for
example, with $\Omega_m = 1$, is also consistent with the cluster
data.

The observed
 cluster mass function is presented in Figure~\ref{fig:eight} together with the
results of CDM
 simulations.  The standard CDM model $(\Omega_m = 1)$ is presented for
several
 bias parameters $b$ 
[where $b$ represents
 the overdensity of galaxies relative to the mass overdensity; 
see \S~\ref{sec:2.1}]; only
the unbiased
 case, $b\sim 1$, is consistent with the COBE results.  This unbiased
model, however, is  excluded by the observed mass function:  it
predicts a
 much larger number of rich clusters than is observed.  The larger
bias
 model of $b\sim 2$, while providing a more
 appropriate mean abundance of rich clusters, is too steep for the observed
mass
 function; it is also incompatible with the COBE results.

The low-density, low-bias CDM model, with $\Omega_m \approx 0.25$ and
$b \approx 1$, (with or without $\Lambda$) is consistent with the 
observed mass function (Fig.~\ref{fig:eight}).
  
How does the CDM model agree with the observations of cluster
correlations?  Bahcall and Cen \cite{bahcall92b}  used the N-body ($N
\sim 10^7$) simulations to determine 
 the model cluster correlation function as a function of cluster mean
 separations $d$. 
The CDM results for clusters corresponding to the
 rich Abell clusters $({\mathrm{richness~class}}\break R \geq 1)$ with $d = 55 
h^{-1}$~Mpc are
 presented in Figure~\ref{fig:nine} together with the observed correlations 
\cite{bahcall83,peacock92}.  The results indicate that the
standard
$\Omega_m = 1$ CDM models are inconsistent with the observations; they cannot
provide
 either the strong amplitude or the large scales ($\gtorder 50 h^{-1}$~Mpc) to 
which the cluster correlations are observed.  Similar results are
found for
 the APM and EDCC clusters.

The low-density, low-bias model is 
 consistent with the observed cluster correlation function.  It 
reproduces both the strong
 amplitude and the large scale to which the cluster correlations are
detected.  Such a model is the only scale-invariant CDM model that is 
consistent
 with both cluster correlations and the cluster mass function.  

The dependence of
 the observed cluster correlation on $d$ was tested in the simulations
(\cite{bahcall92b}).  The results are shown in Figure~\ref{fig:ten} for the
 low-density model.  The dependence of correlation amplitude on mean
separation is
 clearly seen in the simulations.  To compare this result directly
with observations, we plot in Figure~\ref{fig:eleven} the dependence of the 
correlation scale, $r_o$, on $d$ for both the simulations and 
the observations.  The low-density
 model agrees well with the observations, yielding $r_o \approx 0.4 d$, as
observed.  The $\Omega_m = 1$ model, while also showing an increase of $r_o$ 
with
 $d$, yields considerably smaller correlation scales and a much slower
increase of
 $r_o (d)$.

What causes the observed dependence on cluster richness 
[Eqs.~(\ref{eq:two}--\ref{eq:three})]?
The dependence, seen
both
 in the observations and in the simulations, is most likely caused by
 the statistics of rare peak events, which Kaiser \cite{kaiser84} 
suggested as an explanation of the observed strong increase of correlation 
amplitude from
galaxies to
 rich clusters.  The correlation function of rare peaks in a Gaussian
field increases with their selection threshold.  Since more massive
clusters correspond to a
 higher threshold, implying rarer events and thus larger mean
separation, equation~(\ref{eq:three}) results.  A 
 fractal distribution of galaxies and clusters would also produce 
equation~(\ref{eq:three}) (e.g., \cite{szalay85}).

\section{\textsc{Peculiar Motions on Large Scales}}
\label{sec:3}

 How is the mass distributed in the universe?  Does it follow, on the
average,  the light
distribution? To address this important question, peculiar motions on large
scales are studied
 in order to directly trace the mass distribution.  It
 is believed that the peculiar motions (motions relative to a pure
Hubble
 expansion) are caused by the growth of cosmic
structures due to gravity [see the presentations by Ostriker, Peebles, and 
Steinhardt].  
A comparison
 of the mass-density distribution, as reconstructed from peculiar
velocity data, with
 the light distribution (i.e., galaxies) provides information on how
well
 the mass traces light \cite{dekel94,strauss95}. The basic underlying
relation between peculiar velocity and density is given by 
\begin{equation}
\vec\nabla \cdot \vec v = - \Omega_m^{0.6} \delta_m = -\Omega_m^{0.6}
\delta_g/b
\label{eq:five}
\end{equation}
where $\delta_m \equiv (\Delta \rho/\rho)_m$ is the mass overdensity,
$\delta_g$ is the galaxy overdensity, and $b \equiv \delta_g/\delta_m$
is the bias parameter discussed in \S~\ref{sec:2}.
 A formal 
analysis yields a measure of the
 parameter $\beta \equiv \Omega_m^{0.6}/b$. 
Other   methods that place
constraints on $\beta$
 include the anisotropy in the galaxy distribution in the redshift
direction due to peculiar motions (see \cite{strauss95} for a
review).

Measuring peculiar motions is difficult.  The motions are
usually inferred with the aid of measured distances to galaxies or clusters
that are obtained  using some (moderately-reliable) 
distance-indicators (such as
 the Tully-Fisher or $D_n -\sigma$ relations), and 
the measured galaxy redshift.  The peculiar velocity $v_p$ is 
 then determined from the difference between the measured redshift
velocity, $cz$, and
 the measured Hubble velocity, $v_H$, of the system (the latter obtained from 
the
distance-indicator):  $v_p = cz - v_H$.

\begin{figure}[ht]
\plotfiddle{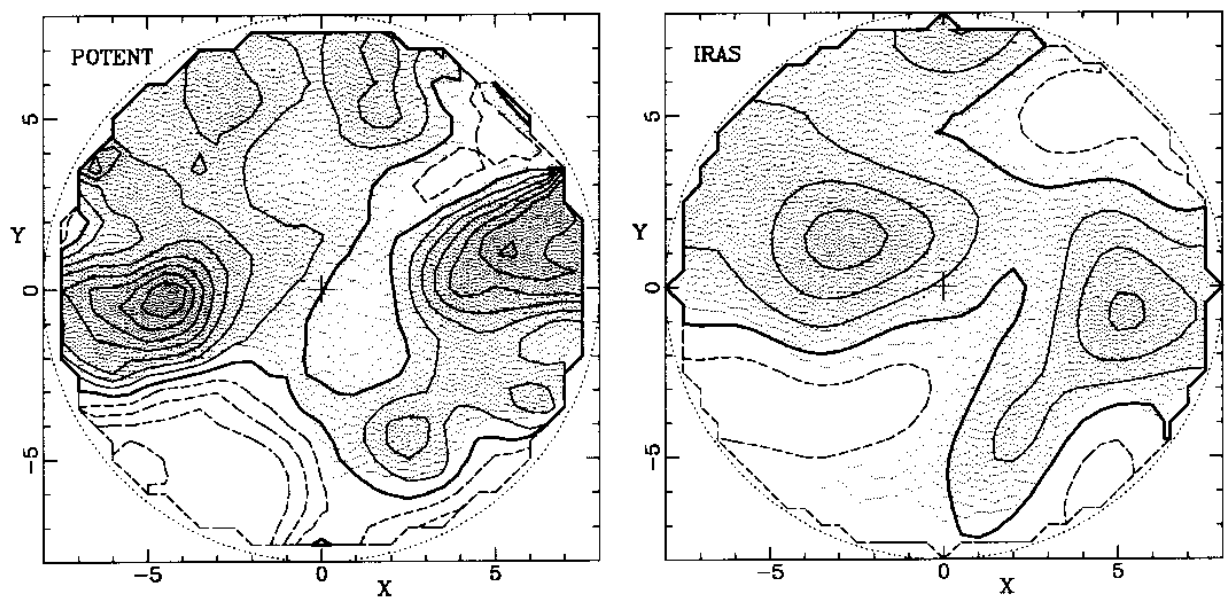}{12cm}{0}{180}{180}{-250}{36}
\caption[]{  A comparison of 
the mass-density (reconstructed from velocity
data) and galaxy-density fields.  The left panel is the density
field
 $\nabla \cdot v$ in the supergalactic plane reconstructed 
from peculiar velocity
data.  The right panel is the independently determined density field of
IRAS galaxies.  The smoothing in both panels is $1200~
{\mathrm{km~ s^{-1}}}$.  The axes
 are labeled in $1000~{\mathrm{km~ s^{-1}}}$.  The Local Group sits at
 the center of each panel (\cite{dekel94,strauss95})\label{fig:twelve}.}
\end{figure}

A comparison between the density distribution of IRAS
 galaxies and the mass-density distribution reconstructed from the
observed peculiar motions
 of galaxies in the supergalactic plane is presented in 
Figure~\ref{fig:twelve} \cite{dekel94,strauss95}.  
The distribution of mass and light in this case appear to be
 similar.

\begin{figure}[ht]
\plotfiddle{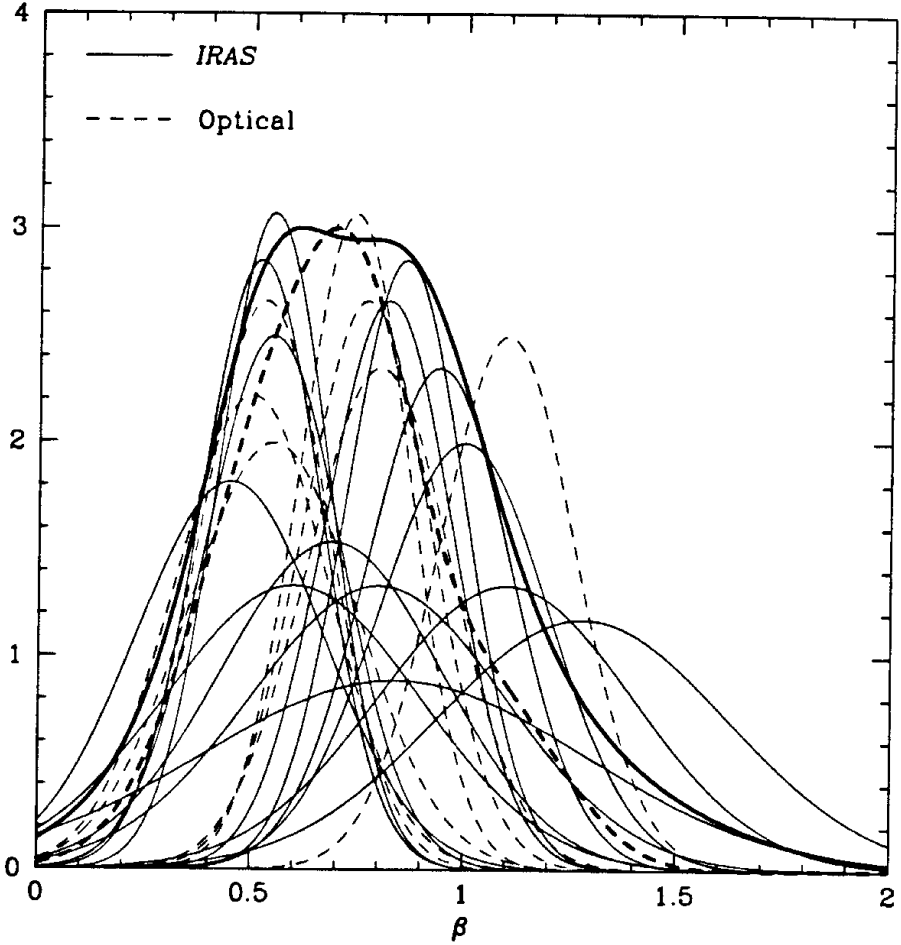}{12cm}{0}{120}{120}{-170}{36}
\caption[]{  Distribution of $\beta = \Omega_m^{0.6}/b$ 
determinations \cite{strauss95}.  Each measurement 
 is shown as a Gaussian of unit integral with mean and standard
 deviations as given by the different observations.  IRAS and optical 
determinations
are plotted with
 different line types.  The (renormalized) sum of all curves (optical
and IRAS
 separately) are shown as the heavy curves\label{fig:thirteen}.}
\end{figure}

A summary of all  measurements of $\beta$ made so far is 
presented in Figure~\ref{fig:thirteen} \cite{strauss95}.  The dispersion
 in the current measurements of $\beta$ is very large; the various
determinations
 range from $\beta \sim 0.4$ to $\sim 1$, implying, for $b \simeq 1, \Omega_m
 \sim 0.2$ to $\sim 1$.  No strong conclusion can therefore be reached at 
present
 regarding the values of $\beta$ or $\Omega_m$.  The larger and more accurate
 surveys currently underway, including high precision velocity
measurements, may lead to the
 determination of $\beta$ and possibly its decomposition into
$\Omega_m$ and $b$ (e.g., \cite{cole94}).

\begin{figure}[ht]
\plotfiddle{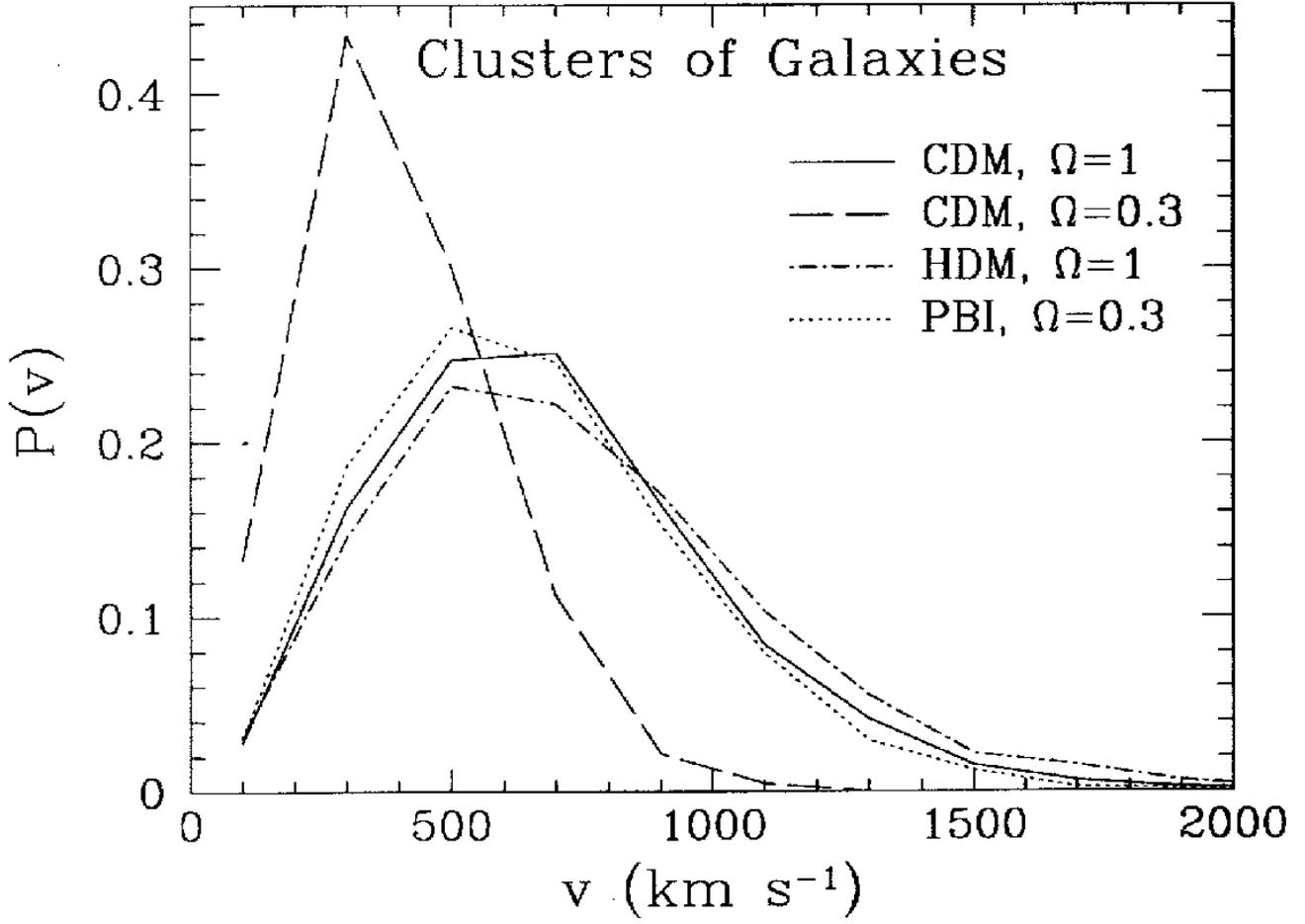}{12cm}{0}{180}{180}{-250}{36}
\caption[]{  The cluster 3-D peculiar 
velocity distribution in four cosmological
models \protect\cite{bahcall94b}\label{fig:fourteen}.}
\end{figure}

\begin{figure}[ht]
\plotfiddle{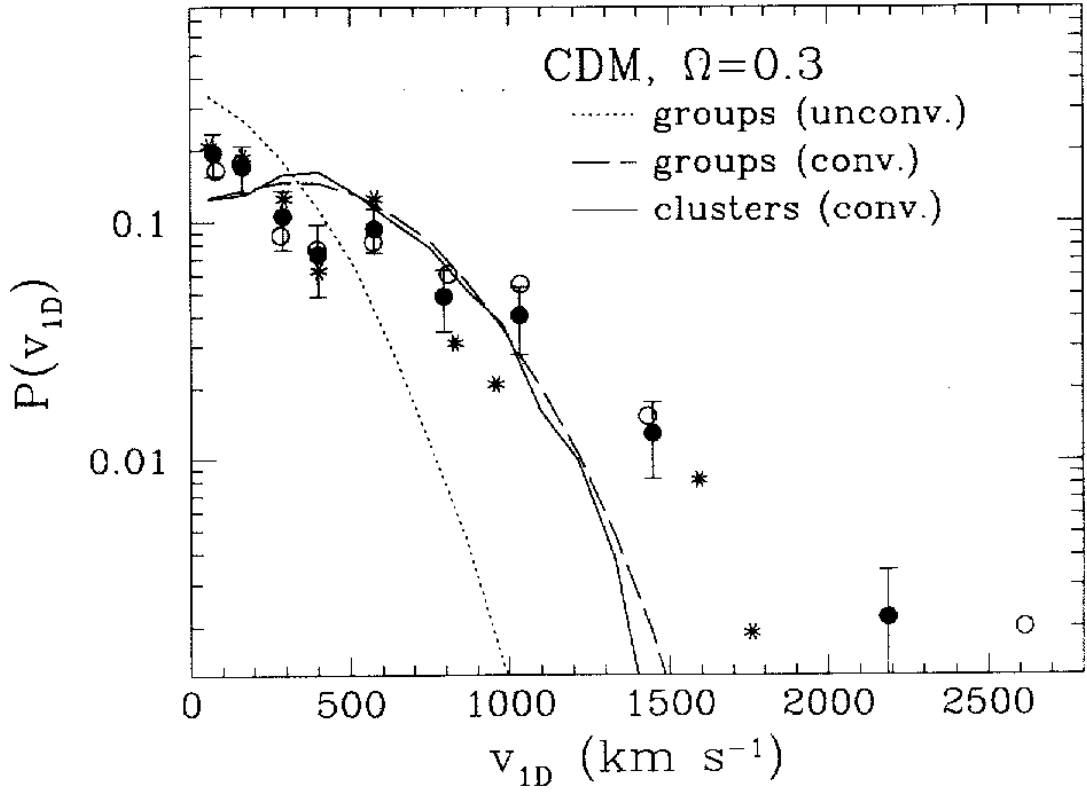}{12cm}{0}{180}{180}{-250}{36}
\caption[]{  Comparison of the currently 
observed 1-D peculiar velocity distribution
of
 groups and clusters of galaxies with expectation from an $\Omega_m = 0.3$
 CDM model.  The dotted line is the model expectation; the dashed and solid 
lines
 are the model convolved with the observed velocity uncertainties
\protect\cite{bahcall94b}\label{fig:fifteen}.}
\end{figure}

Clusters of galaxies can also serve as
 efficient tracers of the large-scale peculiar velocity field in the
universe \cite{bahcall94b}.  Measurements of cluster peculiar velocities are
 likely to be more
 accurate than measurements of individual 
galaxies, since cluster distances can be
determined by averaging
 a large number of cluster members as well as by using different
distance indicators.  Using large-scale cosmological simulations, 
Bahcall et al.~\cite{bahcall94b}
find that clusters
 move reasonably fast in all the cosmological models studied, tracing well the
underlying matter
 velocity field on large scales.  The clusters exhibit a Maxwellian
distribution of
 peculiar velocities as expected from Gaussian initial density
fluctuations.  The model cluster 3-D velocity distribution, 
presented in Figure~\ref{fig:fourteen}, typically peaks at $v \sim
600~{\mathrm{km~ s^{-1}}}$ and extends to high cluster velocities of 
$\sim 2000~{\mathrm{km~ s^{-1}}}$.  The low-density CDM model exhibits 
somewhat lower velocities
(Fig.~\ref{fig:fourteen}).  Approximately 10\% of all 
model rich clusters (1\% for low-density
 CDM) move with $v \gtorder  10^3~{\mathrm{km~ s^{-1}}}$.  A comparison of 
model
 expectation with the available data of cluster velocities is
presented in Figure~\ref{fig:fifteen} 
\cite{bahcall94b}.  The cluster velocity data are not
sufficiently
 accurate at present to place constraints on the models; however,
improved cluster
 velocities, expected in the next several years, should help constrain
the cosmology.

\begin{figure}[ht]
\plotfiddle{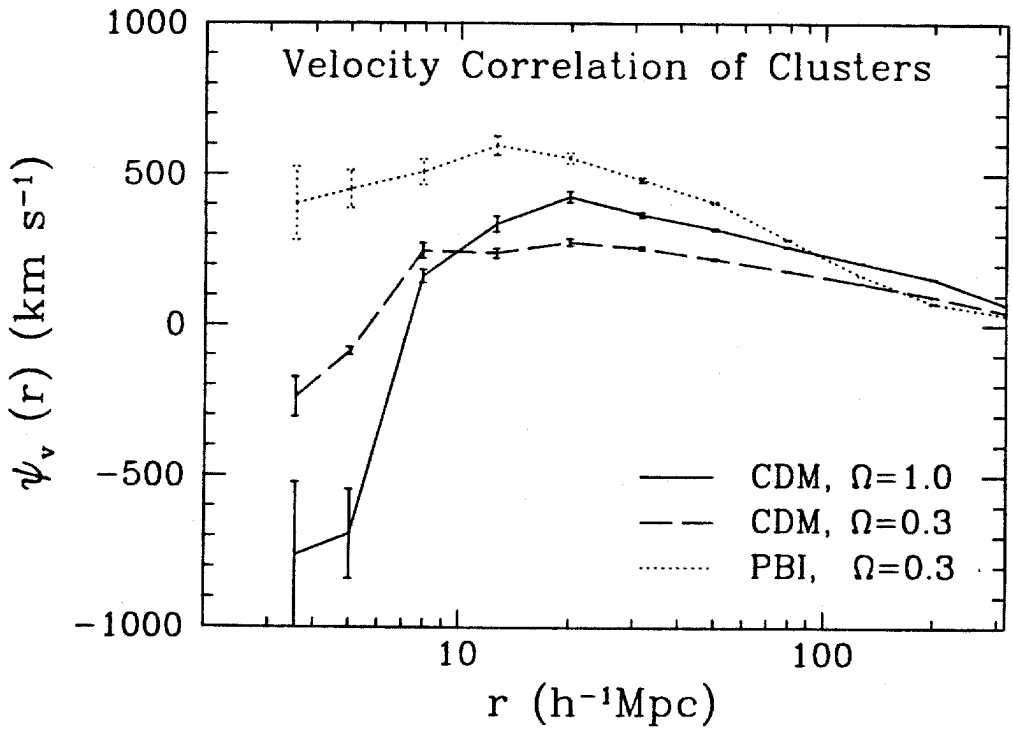}{12cm}{0}{100}{100}{-250}{36}

\caption[]{  The velocity correlation 
function of rich ($R \geq 1$) clusters
 for three models \protect\cite{cen94}\label{fig:sixteen}.}
\end{figure}

Cen, Bahcall and Gramann \cite{cen94} have recently determined the
velocity correlation function
 of clusters in different cosmologies.  They find that close cluster
pairs, with
 separations $r \ltorder 10 h^{-1}$~Mpc, exhibit strong attractive motions; the
pairwise velocities
 depend sensitively on the model.  The mean pairwise attractive
cluster velocities on
 $5 h^{-1}$~Mpc scale ranges from $\sim 1700~{\mathrm{km~ s^{-1}}}$ for 
$\Omega_m = 1$ CDM to $\sim 700~{\mathrm{km~ s^{-1}}}$ for $\Omega_m = 0.3$ CDM
 \cite{cen94}.  The cluster velocity correlation function,
presented in Figure~\ref{fig:sixteen}, is negative on 
small scales---indicating large attractive
velocities, and is
 positive on large scales, to $\sim 200 h^{-1}$~Mpc---indicating significant
bulk motions
 in the models.  None of the models reproduce the very large
 bulk flow of clusters on $150 h^{-1}$~Mpc scale, $v \simeq 689 \pm 178~
{\mathrm{km~ s^{-1}}}$, recently reported by Lauer and Postman \cite{lauer94}. 
 The bulk
 flow expected on this large scale is generally  $\ltorder 200~
{\mathrm{km~ s^{-1}}}$
 for all the models studied ($\Omega_m = 1$ and 
$\Omega_m ~\simeq 0.3$ CDM, and PBI; \cite{cen94,strausscen}).

\section{\textsc{Dark Matter and Baryons in Clusters of Galaxies}}
\label{sec:4}

The
  gravitational potential of clusters of galaxies has
traditionally been studied with the aid of 
 the observed velocity dispersion of the cluster galaxies and the virial
theorem (e.g. \cite{zwicky57,peebles80}).  The cluster potential can also be
estimated
 by the temperature of the hot intracluster gas as determined from
X-ray 
observations (e.g., \cite{jones84,sarazin86}) and by
 the weak gravitational lens distortion of background galaxies caused
by the intervening
 cluster mass \cite{tyson90,kaiser93}.  Sufficient
data are
 now available from optical observations (galaxy velocity dispersion in
clusters), 
 X-ray observations (the temperature of the intracluster gas), and
 initial estimates from gravitational lensing, so that the 
results can be intercompared.  
 The cluster mass determinations are, on the average, consistent with
each other \cite{bahcall95a}.

The optical and X-ray observations of rich clusters of
 galaxies yield cluster masses that range from $\sim 10^{14}$ to $\sim 10^{15} 
h^{-1}
 M_\odot$ within $1.5 h^{-1}$~Mpc radius of the cluster center.  When
 normalized by the cluster luminosity, a median value of $M/L_B \simeq
300 h$ 
is observed for rich clusters.  This mass-to-light ratio implies
 a dynamical mass density of $\Omega_{\mathrm{dyn}} \sim 0.2$ on $\sim 1.5
h^{-1} {\mathrm{~Mpc}}$ scale.  If, as suggested by
 theoretical prejudice, the universe has critical density ($\Omega_m = 1$),
then most of
 the mass in the universe {\it cannot} be concentrated in clusters, groups
 and galaxies; the mass  would have to be distributed 
more
 diffusely than the light.

\begin{figure}[ht]
\plotfiddle{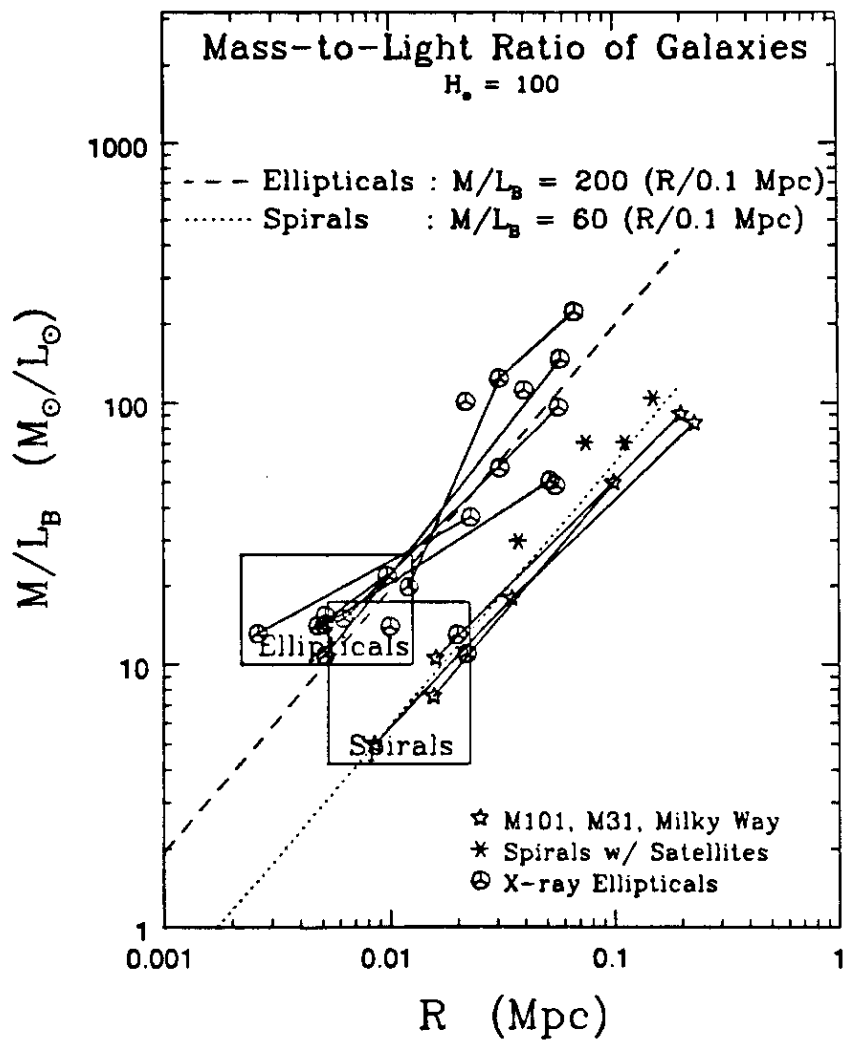}{12cm}{0}{150}{150}{-180}{36}
\caption[]{  The mass-to-light ratio of spiral and 
elliptical galaxies as a function of
 scale.  The large boxes indicate the typical ($\sim 1\sigma$) range of $M/L_B$
 for bright ellipticals and spirals at their luminous (Holmberg)
radii.  ($L_B$ refers
 to total corrected blue luminosity.)  The best-fit $M/L_B \propto R$ lines
 are shown \protect\cite{bahcall95b}\label{fig:seventeen}.}
\end{figure}

\begin{figure}[ht]
\plotfiddle{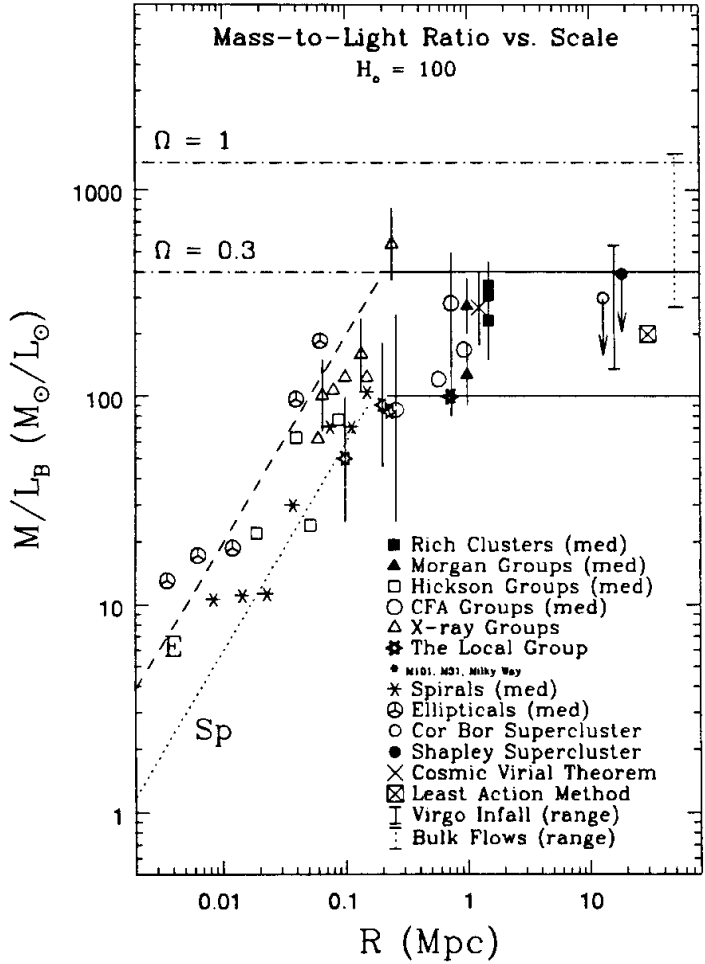}{12cm}{0}{125}{125}{-150}{0}
\caption[]{  A composite mass-to-light ratio 
of different systems---galaxies, groups, clusters, and
superclusters---as a
 function of scale.  The best-fit $M/L_B \propto R$ lines for spirals
 and ellipticals (from Fig.~\ref{fig:seventeen}) are shown.  
Typical $1\sigma$ scatter around
median
 values are shown.  Also presented, for comparison, are the $M/L_B$ (or
 equivalently $\Omega_m$) determinations from the Cosmic Virial Theorem, the
Least Action method, and the range of various reported results from the 
Virgocentric
infall and
 large-scale bulk flows (assuming mass traces light).  The $M/L_B$ 
expected for $\Omega_m = 1$ and $\Omega_m = 0.3$ are indicated 
\protect\cite{bahcall95b}\label{fig:eighteen}.}
\end{figure}

A recent analysis of the mass-to-light
 ratio of galaxies, groups and clusters  \cite{bahcall95b} 
suggests that while the $M/L$ ratio of galaxies increases with scale
 up to radii of $R \sim 0.1$--$0.2 h^{-1}$~Mpc, due
 to the large dark halos around galaxies (see Fig.~\ref{fig:seventeen}), this 
ratio
 appears to flatten and remain approximately constant for groups and
rich clusters, to scales of $\sim 1.5$~Mpc, and possibly even to the 
larger scales of
superclusters (Fig.~\ref{fig:eighteen}).  The flattening occurs at $M/L_B 
\simeq
200$--$300 h$, corresponding
to
 $\Omega_m \sim 0.2$.  This observation may suggest that most of the dark
 matter is associated with the dark halos of galaxies and that
 clusters do {\it not} contain a substantial amount of additional
dark matter, other
 than that associated with (or torn-off from) the galaxy halos, and
 the hot intracluster medium.  Unless the distribution 
 of matter is very different from the distribution of light, with
 large amounts of dark matter in the ``voids" or on very large
 scales, the cluster observations suggest that the mass density in the 
universe may be
 low, $\Omega_m \sim 0.2$ (or $\Omega_m \sim 0.3$ for a small bias
 of $b \sim 1.5$).

Clusters of galaxies contain many baryons.  
Within $1.5 h^{-1}$~Mpc of a rich cluster, the X-ray 
emitting gas contributes $\sim$~3--$10 h^{-1.5}\%$ of
 the cluster virial mass (or $\sim 10$--30\% for $h=1/2$)
\cite{briel92,white95}.  
Visible stars contribute only a small additional amount to this
 value.  Standard Big-Bang nucleosynthesis limits the mean baryon
density of the
 universe to $\Omega_b \sim 0.015 h^{-2}$ \cite{walker91}.  
This suggests that the baryon fraction
 in some rich clusters exceeds that of an $\Omega_m = 1$ universe by
 a large factor \cite{white93,lubin95}.  Detailed
 hydrodynamic simulations \cite{white93,lubin95} suggest that baryons are
not preferentially segregated into
 rich clusters.  It is therefore suggested that either the
mean density
 of the universe is considerably smaller, by a factor of $\sim 3$, than
 the critical density, or that the baryon density of the universe is
 much larger than predicted by nucleosynthesis.  The observed baryonic
mass fraction in
 rich clusters, when combined with the
 nucleosynthesis limit, suggests   
$\Omega_m \sim 0.2$--0.3; this estimate is consistent with
$\Omega_{\mathrm{dyn}} \sim 0.2$ determined from
clusters.

\section{\textsc{Is $\Omega_m < 1$?}}
\label{sec:5}

Much of the observational
 evidence from clusters and from large-scale structure suggests that the
mass density
 of the universe is sub-critical:  $\Omega_m \sim 0.2$--0.3\break 
(\S\S~\ref{sec:2}--\ref{sec:4}).
I summarize these results below:
\begin{itemize}
\item The masses and the $M/L (R)$ relations of galaxies, groups, and clusters 
of 
galaxies suggest $\Omega_m \sim 0.2$--0.3 (\S~\ref{sec:4}).
\item The high baryon fraction in clusters of galaxies suggests 
$\Omega_m \sim 0.2$--0.3 (\S~\ref{sec:4}).
\item Various observations of large-scale structure (the
 mass function, the power spectrum, and the peculiar velocities on Mpc
scale) all suggest $\Omega_m h \sim 0.2$ (for a CDM-type spectrum) 
(\S~\ref{sec:2}).
\item The
 frequent occurrence of evolved systems of galaxies and clusters at
high redshifts
 ($z\sim 1$) also suggest a low-density universe, in which galaxy systems form
 at earlier times.  No quantitative measures are yet available,
however; this topic
 will be further clarified within the next several years using
observations with
 HST, Keck, other deep ground-based observations, and deep X-ray
surveys
 of clusters.
\item If $H_o \sim 70$--80~${\mathrm{km~ s^{-1}~ Mpc^{-1}}}$, as
 indicated by a number of recent observations (\cite{freedman94} and references
therein), then
 the observed age of the oldest stars requires $\Omega_m \ll 1$.
\item Peculiar motions
 on large scales are too uncertain at the present time (suggesting
density
 values that range from $\Omega_m \sim 0.2$ to $\sim 1$) to shed much
 light on $\Omega_m$.  Future results, based on larger and more accurate
surveys, will help constrain this parameter (\S~\ref{sec:3}).
\end{itemize}

\section{\textsc{The SDSS and Large-Scale Structure}}
\label{sec:6}

I will now describe the general characteristics of the Sloan Digital
Sky Survey (SDSS) (in Section \ref{sec:6.1}) and 
then tell you (in Section \ref{sec:6.2})
some of the exciting things we expect to do with this survey using
clusters of galaxies, my own favorite.  You should know that my choice
of topics reflects primarily my own research activity and that the
SDSS will do many extraordinary and important things with individual
galaxies, quasars, and stars.

\subsection{\textit{The Sloan Digital Sky Survey}}
\label{sec:6.1}

Large surveys, especially in three dimensions,
are vital for
 the study of clustering and large-scale structure of the universe.
The
 surveys must cover huge volumes in order to investigate the
largest-scale
 structure.  They can provide accurate, systematic, and complete data
bases of galaxies, clusters, superclusters, voids, and quasars, which can be 
used to
study the
 universal structure and its cosmological implications.  The surveys
will allow both broad
 statistical studies of large samples as well as specific detailed
studies of
 individual systems.

In this section, I summarize some of the fundamental
contributions
 that large-scale surveys will offer for the study of clustering and
 large-scale structure.  I will concentrate on the planned Sloan
Digital Sky
 Survey (SDSS; \cite{york93}), which is the largest survey currently
planned; 
however most of
 the discussion is generally applicable to large-scale imaging and
spectroscopic
 surveys.

The Sloan survey will make a powerful contribution to the
 investigation 
 of large-scale structure because of the survey's size, uniformity,
and the
 high quality of its photometric and spectroscopic data.  These
characteristics will allow
 galaxy clustering in the present-day universe to be measured with
unprecedented
 precision and detail.  The survey will be able to resolve some of
 the unsolved problems listed in \S~\ref{sec:1}, as well as address questions
we
 have not yet thought to ask.  There will be a lot of
 problems for students to solve with these data!

The SDSS will produce
 a complete photometric and spectroscopic survey of half the northern
sky ($\pi$ steradians).  The photometric survey will image the sky in five
colors (${\mathrm{u^\prime,\  g^\prime,\  r^\prime,\  i^\prime,\
z^\prime}}$), using a large array of thirty $2048^2$ pixel
 CCD chips; the imaging survey will contain nearly $5 \times 10^7$
galaxies 
to $g^\prime \sim 23^m$, a comparable number of stars, 
and about $10^6$ quasar candidates
(selected
 on the basis of the five colors).  The imaging data will then
 be used to select the brightest $\sim 10^6$ galaxies and $\sim 10^5$ quasars 
for
 which high-resolution spectra will be obtained (to $r^\prime \sim 18^m$ and
 $19^m$, respectively) using 640 fibers on two high-resolution 
($R=2000$) double-spectrographs.  Both the imaging and spectroscopic surveys
will be done on
 the same 2.5-meter wide-field ($3^\circ$~FOV) special purpose telescope
 located at Apache Point, NM.  The imaging survey will also be used
 to produce a catalog in five colors of all the detected objects
 and their main characteristic parameters.  In the southern
hemisphere, 
the SDSS will image repeatedly a long and narrow strip of the sky, 
$\sim 75^\circ \times 3^\circ$, reaching $2^m$ fainter than 
the large-scale northern survey.
The
 repeated imaging, in addition to allowing the detection of fainter
images, will
 also be crucial for the detection of variable objects.

Figure~\ref{fig:two} shows a  simulated distribution in redshift space 
of galaxies expected from the 
Sloan Survey in a $6^\circ$ thick slice along the survey 
equator. 
This
 figure, based on a cold-dark-matter cosmological model simulation,
 corresponds to approximately
 6\% of the galaxy redshift survey.  The redshift histogram of galaxies
  in
 the simulated spectroscopic survey peaks at $z \sim 0.1$--0.15, with
 a long tail to $z \sim 0.4$.

Several thousand rich clusters of
 galaxies will be identified in the survey, many with a large number
 of measured galaxy redshifts per cluster (from the complete
spectroscopic survey).  Several
 hundred large superclusters will also be identified.  This will
provide at least
 a ten-fold increase in the number of clusters and superclusters over
  presently available samples. 

Currently under construction, the test year for the SDSS system
 will begin observations in 1997.  More detailed information on the SDSS 
project
is provided in \cite{york93}.

\subsection{\textit{Clusters of Galaxies}}
\label{sec:6.2}

Large-scale surveys such as the SDSS will provide a much
 needed advance in the systematic study of clusters of galaxies, which
is
 currently limited by the unavailability of modern, accurate,
complete, and objectively-selected
 catalogs of clusters, and by the limited photometric and redshift
information for
 the catalogs that do exist.  The SDSS will select clusters
algorithmically.  From
 the $10^6$ galaxies in the spectroscopic sample, we can identify
clusters by
 searching for density enhancements in redshift space.  We can also
identify clusters
 from the much larger photometric sample ($5 \times 10^7$ galaxies) by 
searching
for density
 enhancements in position-magnitude-color space.  We expect to find
approximately 4000
 clusters of galaxies with a redshift tail to $z \sim 0.5$; many
 of the clusters will have a large number of measured galaxy redshifts
 per cluster (hundreds of redshifts for nearby clusters), and at least
one
 or two redshifts for the more distant clusters.  This 
complete
 sample of clusters can be used to investigate numerous topics in the
 study of clustering and large-scale structure.  I briefly illustrate
some of
 these topics below.  The solutions of the problems
outlined
 below will involve the participation of many graduate
students!

\subsubsection{Clusters as Tracers of Large-Scale Structure}

Clusters are efficient tracers
 of the large-scale structure.  Studies of the cluster
distribution have
 yielded important results even though the number of cluster redshifts
in complete
 samples has been quite small, $\sim 100$ to 300.  The SDSS will
 put such studies on a much firmer basis because of the increased 
 sample size (roughly an order of magnitude more redshifts), the
objective and automated identification
 methods, and the use of cluster-finding algorithms that are less
subject
 to projection contamination.

Catalogs of nearby superclusters have been constructed using
Abell
 clusters with known redshifts, revealing structures on scales as
large as $\sim 150 h^{-1}$~Mpc \cite{bahcall84}.  There are claims for still
larger, $\sim 300 h^{-1}$~Mpc structures in the Abell cluster distribution 
\cite{tully87}; these
 are still controversial \cite{postman89}.  The SDSS redshift
survey will
 identify a considerably bigger sample of superclusters and will clarify the
possible existence of structure on very large scales.  The sizeable 
cluster sample should also clarify the relation between
 the cluster distribution and the galaxy distribution.  
 Cluster peculiar velocities, obtained directly from distance 
indicators such as the $D_n - \sigma$
measurements and
 indirectly from the anisotropy of the cluster correlation function in
redshift space, will provide information about the internal dynamics of
superclusters and the mass
 distribution on large scales.

The high amplitude of the cluster correlation
 function, $\xi_{cc}(r)$, provided early evidence
 for strong clustering on large scales (\S~\ref{sec:2}).  
The complete sample of SDSS clusters, objectively defined from uniform
and accurate
 photometric and spectroscopic data, will allow a  measurement of  $\xi_{cc}$ 
 free from systematic effects.  We can extend the correlation analysis
in several
 ways, examining in detail the richness dependence of $\xi_{cc}$ as
well
 as the dependence of the cluster correlation on other cluster
properties (e.g., velocity dispersion, morphology).

The alignment of galaxies with their host clusters, and of clusters
with
 their neighboring clusters, their host superclusters, and other
large-scale structures such
 as sheets and filaments is of great interest.  While the alignment 
of cD galaxies with
their
 clusters is relatively well established \cite{binggeli82}, alignments
at larger scales are
 still uncertain \cite{chincarini88}.  The SDSS will provide a
very
 large sample for alignment studies, including accurate position
angles and inclinations.  The
 existence of large-scale alignments may place interesting constraints
on the origin
 of structure in various cosmological models \cite{west89}.

\subsubsection{Global Cluster Properties}

The complete cluster survey will allow a detailed
investigation of intrinsic
 cluster properties.  From the 2-D and 3-D information, we will
 be able to determine accurately such properties as cluster richness,
morphology, density
 and density profile, core radius, velocity dispersion profile,
optical luminosity, and galaxy
 content, and to look for correlations between these properties.  The
cluster catalogs will
 be matched with data in other bands, in particular the X-ray.  
This will allow a systematic detailed study of global cluster
properties and
 their cross-correlation.\break  Measurements of galaxy density, 
morphology, velocity dispersion, and
 X-ray emission will shed light on the nature of the intracluster
 medium and its impact on the member galaxies, and on the relative
 distribution of baryonic and dark matter.  High surface mass
density clusters
 will be candidates for gravitational lenses; we can search the
photometric data
 for systematically distorted background galaxies, especially in the
 deeply-observed southern strip, and target these
 clusters for yet deeper imaging surveys with larger telescopes.
These lensing studies
 will allow mapping out the dark matter distribution within the
clusters \cite{tyson90,kaiser93}.

The large statistical sample of clusters
 will produce new insights into a number of issues associated with
structure
 formation and evolution:
\begin{itemize}
\item  We will be able to calculate velocity dispersions for
 $\sim 1000$ clusters with $z\ltorder  0.2$ compared to a few dozen clusters
 with reliable dispersion measurements currently available.  For
nearer clusters, we will have
 well-sampled galaxy density profiles and velocity dispersion
profiles; these allow careful
 cluster mass determinations.  The distributions of cluster masses and
velocity dispersions, i.e., the cluster mass function and velocity function 
will be
determined accurately. 
\item  With optical luminosities, galaxy profiles, and velocity
dispersion profiles, we
 can measure accurate mass-to-light ratios for a large sample of
 clusters.  Combined with X-ray observations of clusters, the baryon
fraction can
 be determined more accurately.  These provide constraints on $\Omega_m$ and
the bias
 parameter of galaxies in clusters.
\item  The density and dispersion profiles also
 retain clues about the history of cluster formation.  The frequency
of subclustering
 and non-virial structures in clusters \cite{geller90} tells us whether
clusters
 are dynamically old or young.  In a gravitational instability model
with $\Omega_m = 1$, structure continues to form today, 
while in a low-$\Omega$, open
 universe, clustering freezes out at moderate redshift.
Cluster profiles and
 substructure statistics thus provide a diagnostic for $\Omega_m$, a
diagnostic  which is
independent of 
 the direct measures of the mass density.  
\item  The evolution of the
 cluster population provides 
 another diagnostic for $\Omega_m$.  Current
investigations are limited by
 the small numbers of known high-\break redshift clusters, and by the fact
 that high- and low-redshift clusters are\break selected in different ways.
The
 uniform SDSS cluster sample will greatly improve the current
situation.  The number
 of clusters will be large; we will have redshifts for brightest
cluster
 galaxies to $z \approx 0.5$ by extending the main redshift survey
limit $\sim 1^m$ fainter for brightest cluster galaxy candidates, 
and for more distant clusters we can
 obtain fairly accurate estimated redshifts from the photometry alone
(using the colors
 and apparent luminosity function of member galaxies).  The southern
photometric survey will
 probe the cluster population to redshifts above unity.
\end{itemize}

\subsubsection{Large-Scale Motions of Clusters}

Peculiar motions of clusters of galaxies can be
determined from the
 SDSS survey using  distance indicators such as Tully-Fisher or $D_n -
\sigma$ 
for cluster
 galaxies, the cluster luminosity function, 
or the Brightest-Cluster-Member indicator.  This will allow
a determination
 of the large-scale peculiar motion of clusters to hundreds of Mpc.

\section{\textsc{Summary}}
\label{sec:7}

Observations of large-scale structure provide a powerful tool
 for 
 determining the structure of our universe, for understanding galaxy and
structure formation, and for constraining cosmological models.  While
extraordinary 
progress has been\break made over the
 last decade, many important unsolved problems remain.  The large
 observational surveys currently
 underway provide a way of resolving some of these questions [see also
the contribution by P. Steinhardt, these proceedings]. The interested
student has a great opportunity to participate in fundamental discoveries.

What problems can we expect to solve in the next
 decade?  I believe that the large redshift surveys of galaxies,
clusters of
 galaxies, and quasars will provide a quantitative description of 
 structure on a wide range of scales, from $\sim 1$ to 
$\sim 10^3 h^{-1}$~Mpc, including the determinations of the power spectrum, the
correlation function, the topology, and other improved statistics 
(\S\S~\ref{sec:2},\ref{sec:6}).  The dependence
of these
 statistical measures on galaxy type, luminosity,
surface brightness, and system
 type (galaxies, clusters, quasars) will also be determined 
and used to constrain cosmological
models of
 galaxy formation.  Accurate measurements on large scales  of the 
peculiar motions of galaxies
and  clusters,  and  the mass distribution on large scales may take longer.  
 Detailed X-ray surveys of galaxy clusters, combined with the
large
 redshift and lensing surveys,  will be able to resolve the
 problem of the baryon fraction in clusters and test suggested
explanations (\S~\ref{sec:4}).  Deep optical and X-ray surveys, 
using HST, Keck, ROSAT, ASCA, AXAF, SIRTF, plus new microwave
background studies, and other deep 
ground-based and space-based observations, should
allow a
 considerable increase in our understanding of the time evolution of
galaxies, clusters
 of galaxies, quasars, and large-scale structure.  Since the computed evolution
of galaxy
 systems depends strongly on the assumed cosmology and especially on the
mass density, with low-density models evolving at much earlier times than high
density
 models, observations of the kind described here 
will provide some of the most critical
clues needed to constrain the cosmological model over the next decade
 (\S\S~\ref{sec:5},\ref{sec:6}).

The quantitative description of large-scale structure and its
 evolution, when compared with state-of-the-art cosmological simulations to
be available
 in the next decade (e.g. J. P. Ostriker, these proceedings), will enhance our
 chances of determining the cosmology of our universe.  Will this goal
be
 reached within the next decade?  We will have to meet again in
 a decade and see.

\section*{\textsc{Acknowledgments}}

It is a pleasure to thank J. N. Bahcall, D. Eisenstein, K. Fisher, J.
P. Ostriker, P. J. E. Peebles, H. Rood, D. Spergel, and M. Strauss for
stimulating discussions and helpful comments on the manuscript. 
The work by N. Bahcall and collaborators
 is  supported by NSF grant AST93-15368 and NASA graduate training
grant NGT-51295.

\section*{\textsc{Bibliographic Notes}}
\markright{\textsc{Bibliographic Notes}}

\begin{itemize}

\item {\bf Bahcall, N. A. 1988, ARA\&A, 26, 631.} A detailed review of
the large-scale structure of the universe as traced with clusters of
galaxies.
\item {\bf Bahcall, N. A., \& Soneira, R. M. 1983, ApJ, 270, 20.} The
first determination of the three-dimensional cluster correlation
function and the richness dependence of the cluster correlations.
\item {\bf Geller, M., \& Huchra, J. 1989, Science, 246, 897.} A clear
and exciting 
summary of the large-scale structure observed in the CFA redshift
survey.
\item {\bf Maddox, S., Efstathiou, G., Sutherland, W., \& Loveday, J. 
1990,\break
MNRAS, 242, 43p.} A recent determination of the angular correlation
function of galaxies using the APM survey.
\item {\bf Peebles, P. J. E. 1980, The Large-Scale Structure of the
Universe; 1993, Principles of Physical Cosmology (Princeton: Princeton 
University\break 
Press).} The standard monographs from which students (and professors) have
learned the subjects of large-scale structure and cosmology.
\item {\bf Strauss, M., \& Willick, J. 1995, Physics Reports, 261, 271.} A
comprehensive review of the density and peculiar velocity fields of
nearby galaxies.
\item {\bf York, D., Gunn, J.~E., Kron, R., Bahcall, N.~A., Bahcall, J.~N., 
Feldman,
P.~D., Kent, S.~M., Knapp, G.~R., Schneider, D., \& Szalay, A. 1993, 
A Digital Sky Survey of the Northern Galactic Cap, NSF
proposal.} A detailed summary of the planned Sloan Digital Sky Survey.
\end{itemize}

\end{document}